\documentclass[12pt]{article}
\usepackage{times}

\usepackage{amsmath}
\usepackage{amssymb}

\usepackage{graphicx}

\usepackage{color} 

\usepackage{setspace}
\usepackage{xspace}
\def\MixMapper{\emph{MixMapper}\xspace}
\def\TreeMix{\emph{TreeMix}\xspace}

\usepackage[labelfont=bf,labelsep=period,justification=raggedright,singlelinecheck=false]{caption}
\usepackage{float}

\usepackage{url}

\usepackage{NAR-natbib}
\makeatletter
\renewcommand{\@biblabel}[1]{\quad#1.}
\makeatother

\date{}

\title{Efficient moment-based inference of admixture parameters and sources of gene flow}

\author
{Mark Lipson,$^{1\#}$ Po-Ru Loh,$^{1\#}$ Alex Levin,$^{1}$ \\
 David Reich,$^{2,3}$ Nick Patterson,$^{2}$ Bonnie Berger$^{1,2,\ast}$ \\
\\
\normalsize{$^{1}$Department of Mathematics and Computer Science and
  Artificial Intelligence Laboratory,}\\
\normalsize{Massachusetts Institute of Technology, Cambridge, MA 02139}\\
\normalsize{$^{2}$Broad Institute, Cambridge, MA 02142}\\
\normalsize{$^{3}$Department of Genetics, Harvard Medical School, Boston, MA 02115}\\
\\
\normalsize{$^\#$These authors contributed equally to this work.}\\
\normalsize{$^\ast$To whom correspondence should be addressed:}\\
\normalsize{Department of Mathematics 2-373}\\
\normalsize{Massachusetts Institute of Technology}\\
\normalsize{77 Massachusetts Avenue, Cambridge, MA 02139}\\
\normalsize{Tel: (617) 253-1827; Fax: (617) 258-5429}\\
\normalsize{E-mail: \texttt{bab@mit.edu}}\\
\\
}

\date{}

\begin{document}
\maketitle
\newpage

\section*{Abstract}
The recent explosion in available genetic data has led to significant advances in understanding the demographic histories of and relationships among human populations.  It is still a challenge, however, to infer reliable parameter values for complicated models involving many populations.  Here we present \MixMapper, an efficient, interactive method for constructing phylogenetic trees including admixture events using single nucleotide polymorphism (SNP) genotype data.  \MixMapper implements a novel two-phase approach to admixture inference using moment statistics, first building an unadmixed scaffold tree and then adding admixed populations by solving systems of equations that express allele frequency divergences in terms of mixture parameters.  Importantly, all features of the model, including topology, sources of gene flow, branch lengths, and mixture proportions, are optimized automatically from the data and include estimates of statistical uncertainty.  \MixMapper also uses a new method to express branch lengths in easily interpretable drift units.  We apply \MixMapper to recently published data for HGDP individuals genotyped on a SNP array designed especially for use in population genetics studies, obtaining confident results for 30 populations, 20 of them admixed.  Notably, we confirm a signal of ancient admixture in European populations---including previously undetected admixture in Sardinians and Basques---involving a proportion of 20--40\% ancient northern Eurasian ancestry.
\newpage

\section*{Introduction}

The most basic way to represent the evolutionary history of a set of species or populations is through a phylogenetic tree, a model that in its strict sense assumes that there is no gene flow between populations after they have diverged~\citep{cavalli1967phylogenetic}.  In many settings, however, groups that have split from one another can still exchange genetic material. %
This is certainly the case for human population history, during the course of which populations have often diverged only incompletely or diverged and subsequently mixed again~\citep{India, wall2009detecting, laval2010formulating, green2010draft, reich2010genetic, gravel2011demographic, draft7}.  To capture these more complicated relationships, previous studies have considered models allowing for continuous migration among populations~\citep{wall2009detecting, laval2010formulating, gravel2011demographic} or have extended simple phylogenetic trees into \emph{admixture trees}, in which populations on separate branches are allowed to re-merge and form an admixed offspring population~\citep{chikhi2001estimation, MLadm, India, sousa2009approximate, draft7}.  Both of these frameworks, of course, still represent substantial simplifications of true population histories, but they can help capture a range of new and interesting phenomena.

Several approaches have previously been used to build phylogenetic trees incorporating admixture events from genetic data.
First, likelihood methods~\citep{chikhi2001estimation, MLadm, sousa2009approximate} use a full probabilistic evolutionary model, which allows a high level of precision with the disadvantage of greatly increased computational cost.  Consequently, likelihood methods can in practice only accommodate a small number of populations~\citep{wall2009detecting, laval2010formulating, gravel2011demographic, siren2011reconstructing}.  Moreover, the tree topology must generally be specified in advance, meaning that only parameter values can be inferred automatically and not the arrangement of populations in the tree.  By contrast, the moment-based methods of \citet{India} and \citet{draft7} use only means and variances of allele frequency divergences.  Moments are simpler conceptually and especially computationally, and they allow for more flexibility in model conditions.  Their disadvantages can include reduced statistical power and difficulties in designing precise estimators with desirable statistical properties (e.g., unbiasedness)~\citep{MLadm}. %
Finally, a number of studies have considered ``phylogenetic networks,'' which generalize trees to include cycles and multiple edges between pairs of nodes and can be used to model population histories involving hybridization~\citep{huson2006application,yu2012probability}.  However, these methods also tend to be computationally expensive.%

In this work, we introduce \MixMapper, a new computational tool that fits admixture trees by solving systems of moment equations involving the pairwise distance statistic $f_2$~\citep{India,draft7}, which is the average squared allele frequency difference between two populations.  The theoretical expectation of $f_2$ can be calculated in terms of branch lengths and mixture fractions of an admixture tree and then compared to empirical data.  \MixMapper can be thought of as a generalization of the \emph{qpgraph} package~\citep{draft7}, which takes as input genotype data, along with a proposed arrangement of admixed and unadmixed populations, and returns branch lengths and mixture fractions that produce the best fit to allele frequency moment statistics measured on the data.  \MixMapper, by contrast, performs the fitting in two stages, first constructing an unadmixed scaffold tree via neighbor-joining and then automatically optimizing the placement of admixed populations onto this initial tree.  Thus, no topological relationships among populations need to be specified in advance.  

Our method is similar in spirit to the independently developed
\TreeMix package~\citep{pickrell2012inference}.  Like \MixMapper,
\TreeMix builds admixture trees from second moments of allele
frequency divergences, although it does so via a composite likelihood
maximization approach made tractable with a multivariate normal
approximation.  Procedurally, \TreeMix initially fits a full set of populations as an unadmixed tree, and gene flow edges are added sequentially
to account for the greatest errors in the
fit~\citep{pickrell2012inference}.  This format makes \TreeMix
well-suited to handling very large trees: the entire fitting process
is automated and can include arbitrarily many admixture events
simultaneously.  In contrast, \MixMapper begins with a carefully screened unadmixed scaffold tree
to which admixed populations are added with best-fitting parameter values, an interactive design that enables precise modeling of particular populations of interest.

We use \MixMapper to model the ancestral relationships among 52
populations from the CEPH-Human Genome Diversity Cell Line Panel
(HGDP)~\citep{rosenberg2002genetic, li2008worldwide} using recently
published data from a new, specially ascertained SNP array designed
for population genetics
applications~\citep{keinan2007measurement,draft7}.  Previous studies of
these populations have built simple phylogenetic
trees~\citep{li2008worldwide, siren2011reconstructing}, identified a
substantial number of admixed populations with likely
ancestors~\citep{draft7}, and constructed a large-scale admixture
tree~\citep{pickrell2012inference}.  Here, we add an additional level
of quantitative detail, obtaining best-fit admixture parameters with
bootstrap error estimates for 30 HGDP populations, of which 20 are
admixed. The results include, most notably, a significant admixture
event~\citep{draft7} in the history of all sampled European
populations, among them Sardinians and Basques.

\section*{New Approaches}

The central problem we consider is: given an array of
SNP data sampled from a set of individuals grouped by
population, what can we infer about the admixture
histories of these populations using simple statistics that are functions of their allele frequencies?  
Methodologically, the %
\MixMapper workflow (Figure~\ref{fig:flow_chart}) proceeds as follows.  We begin by computing $f_2$ distances between all pairs of study populations, from which we construct an unadmixed phylogenetic subtree to serve as a scaffold for subsequent mixture fitting.  The choice of populations for the scaffold is done via initial filtering of populations that are clearly admixed according to the 3-population test~\citep{India, draft7}, followed by selection of a subtree that is approximately additive along its branches, as is expected in the absence of admixture (see Material and Methods and Text~\ref{text:f-stats+admixture} for full details).

Next, we expand the model to incorporate admixtures by attempting to fit each population not in the scaffold as a mixture between some pair of branches of the scaffold.  Putative admixtures imply algebraic relations among $f_2$ statistics, which we test for consistency with the data, allowing us to identify likely sources of gene flow and estimate mixture parameters (Figure~\ref{fig:table_key}; Text~\ref{text:f-stats+admixture}).  After determining likely two-way admixture events, we further attempt to fit remaining populations as three-way mixtures involving the inferred two-way mixed populations, by similar means.  Finally, we use a new formula to convert the $f_2$ tree distances into absolute drift units (Text~\ref{text:het+drift}).  %
Importantly, we apply a bootstrap resampling scheme~\citep{efron1979bootstrap, efron1986bootstrap} to obtain ensembles of predictions, rather than single values, for all model variables.  This procedure allows us to determine confidence intervals for parameter estimates and guard against overfitting.  
For a data set on the scale of the HGDP, after initial setup time on the order of an hour, \MixMapper determines the best-fit admixture model for a chosen population in a few seconds, enabling real-time interactive investigation.

\section*{Results}

\subsection*{Simulations}

To test the inference capabilities of \MixMapper on populations with known histories, we ran it on two data sets generated with the coalescent simulator \texttt{ms}~\citep{hudson2002generating} and designed to have similar parameters to our human data.  In both cases, we simulated 500 regions of 500 kb each for 25 diploid individuals per population, with an effective population size of 5,000 or 10,000 per population, a mutation rate of $0.5 \times 10^{-8}$ per base per generation (intentionally low so as not to create unreasonably many SNPs), and a recombination rate of $10^{-8}$ per base per generation.  Full \texttt{ms} commands can be found in Material and Methods. We ascertained SNPs present at minor allele frequency 0.05 or greater in an outgroup population and then removed that population from the analysis.

\begin{figure}[H]
\begin{center}
\includegraphics{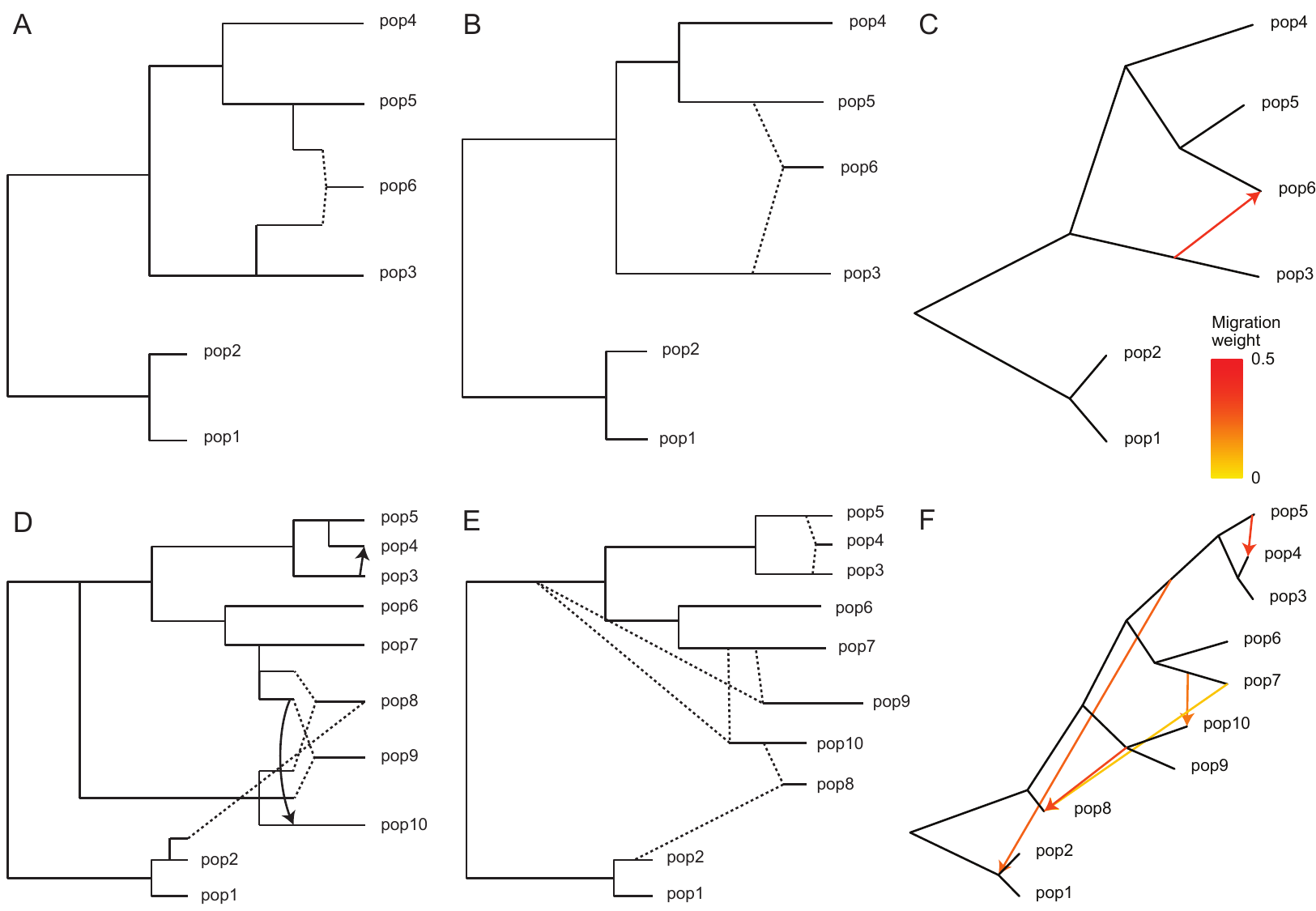}
\end{center}
\caption{{\bf Results with simulated data.} (A-C) First simulated admixture tree, with one admixed population.  Shown are: (A) the true phylogeny, (B) \MixMapper results, and (C) \TreeMix results.  (D-F) Second simulated admixture tree, with four admixed populations.  Shown are: (D) the true phylogeny, (E) \MixMapper results, and (F) \TreeMix results.  In (A) and (D), dotted lines indicate instantaneous admixtures, while arrows denote continuous (unidirectional) gene flow over 40 generations.  Both \MixMapper and \TreeMix infer point admixtures, depicted with dotted lines in (B) and (E) and colored arrows in (C) and (F).  In (B) and (E), the terminal drift edges shown for admixed populations represent half the total mixed drift.  Full inferred parameters from \MixMapper are given in Table~\ref{tab:sim_details}.}
\label{fig:simulations}
\end{figure}

For the first admixture tree, we simulated six non-outgroup populations, with one of them, pop6, admixed (Figure~\ref{fig:simulations}A).  Applying \MixMapper, no admixtures were detected with the 3-population test, but the most additive subset with at least five populations excluded pop6 (max deviation from additivity $2.0 \times 10^{-4}$ versus second-best $7.7 \times 10^{-4}$; see Material and Methods), so we used this subset as the scaffold tree.  We then fit pop6 as admixed, and \MixMapper recovered the correct gene flow topology with 100\% confidence and inferred the other parameters of the model quite accurately (Figure~\ref{fig:simulations}B; Table~\ref{tab:sim_details}).  For comparison, we also analyzed the same data with \TreeMix and again obtained accurate results (Figure~\ref{fig:simulations}C).

For the second test, we simulated a complex admixture scenario involving 10 non-outgroup populations, with six unadmixed and four admixed (Figure~\ref{fig:simulations}D).  In this example, pop4 is recently admixed between pop3 and pop5, but over a continuous period of 40 generations.  Meanwhile, pop8, pop9, and pop10 are all descended from older admixture events, which are similar but with small variations (lower mixture fraction in pop9, 40-generation continuous gene flow in pop10, and subsequent pop2-related admixture into pop8).  In the first phase of \MixMapper, the recently admixed pop4 and pop8 were detected with the 3-population test.  From among the other eight populations, a scaffold tree consisting of pop1, pop2, pop3, pop5, pop6, and pop7 provided thorough coverage of the data set and was more additive (max deviation $3.5 \times 10^{-4}$) than the secon-best six-population scaffold ($5.4 \times 10^{-4}$) and the best seven-population scaffold ($1.2 \times 10^{-3}$).  Using this scaffold, \MixMapper returned very accurate and high-confidence fits for the remaining populations (Figure~\ref{fig:simulations}E; Table~\ref{tab:sim_details}), with the correct gene flow topologies inferred with 100\% confidence for pop4 and pop10, 98\% confidence for pop9, and 61\% confidence for pop8 (fit as a three-way admixture; 39\% of replicates placed the third gene flow source on the branch adjacent to pop2, as shown in Table~\ref{tab:sim_details}).  In contrast, \TreeMix inferred a less accurate admixture model for this data set (Figure~\ref{fig:simulations}F). \TreeMix correctly identified pop4 as admixed, and it placed three migration edges among pop7, pop8, pop9, and pop10, but two of the five total admixtures (those originating from the common ancestor of pops 3--5 and the common ancestor of pops 9--10) did not correspond to true events.  Also, \TreeMix did not detect the presence of admixture in pop9 or the pop2-related admixture in pop8.

\subsection*{Application of \MixMapper to HGDP data}

Despite the focus of the HGDP on isolated populations, most of its 53 groups exhibit signs of admixture detectable by the 3-population test, as has been noted previously~\citep{draft7}.  Thus we hypothesized that applying \MixMapper to this data set would yield significant insights.  Ultimately, we were able to obtain comprehensive results for 20 admixed HGDP populations (Figure~\ref{fig:mix_tree}), discussed in detail in the following sections.  

\subsection*{Selection of a 10-population unadmixed scaffold tree}

To construct an unadmixed scaffold tree for the HGDP data to use in fitting admixtures, we initially filtered the list of 52 populations (having removed San due to ascertainment of our SNP panel in a San individual; see Material and Methods) with the 3-population test, leaving only 20 that are potentially unadmixed. %
We further excluded Mbuti and Biaka Pygmies, Kalash, Melanesian, and Colombian from the list of candidate populations due to external evidence of admixture~\citep{loh2013inferring}.

It is desirable to include a wide range of populations in the unadmixed scaffold tree to provide both geographic coverage and additional constraints that facilitate the fitting of admixed populations (see Material and Methods).  Additionally, incorporating at least four continental groups provides a fairer evaluation of additivity, which is roughly equivalent to measuring discrepancies in fitting phylogenies to quartets of populations.  If all populations fall into three or fewer tight clades, however, any quartet must contain at least two populations that are closely related.  At the same time, including too many populations can compromise the accuracy of the scaffold. We required that our scaffold tree include representatives of at least four of the five major continental groups in the HGDP data set (Africa, Europe, Oceania, Asia, and the Americas), with at least two populations per group (when available) to clarify the placement of admixing populations and improve the geographical balance.  Subject to these conditions, we selected an approximately unadmixed scaffold tree containing 10 populations, which we found to provide a good balance between additivity and comprehensiveness: Yoruba, Mandenka, Papuan, Dai, Lahu, Japanese, Yi, Naxi, Karitiana, and Suru\'{i} (Figure~\ref{fig:mix_tree}B). 
These populations constitute the second-most additive (max deviation $1.12 \times 10^{-3}$) of 21 similar trees differing only in which East Asian populations are included (range $1.12$--$1.23 \times 10^{-3}$); we chose them over the most-additive tree because they provide slightly better coverage of Asia.
To confirm that modeling these 10 populations as unadmixed in \MixMapper is sensible, we checked that none of them can be fit in a reasonable way as an admixture on a tree built with the other nine (see Material and Methods).  Furthermore, we repeated all of the analyses to follow using nine-population subsets of the unadmixed tree as well as an alternative 11-population tree and confirmed that our results are robust to the choice of scaffold (Figures~\ref{fig:alt_scaffold}--\ref{fig:alt_alphas}; Tables~\ref{tab:european_mixes_alt1}--\ref{tab:double_mixes_alt1}).

\subsection*{Ancient admixture in the history of present-day European populations}

A notable feature of our unadmixed scaffold tree is that it does not contain any European populations. \citet{draft7} previously observed negative $f_3$ values indicating admixture in all HGDP Europeans other than Sardinian and Basque.
Our \MixMapper analysis uncovered the additional observation that potential trees containing Sardinian or Basque along with representatives of at least three other continents are noticeably less additive than four-continent trees of the same size without Europeans: from our set of 15 potentially unadmixed populations, %
none of the 100 most additive 10-population subtrees include Europeans.  This points to the presence of admixture in Sardinian and Basque as well as the other European populations.

\begin{figure}[H]
\begin{center}
\includegraphics{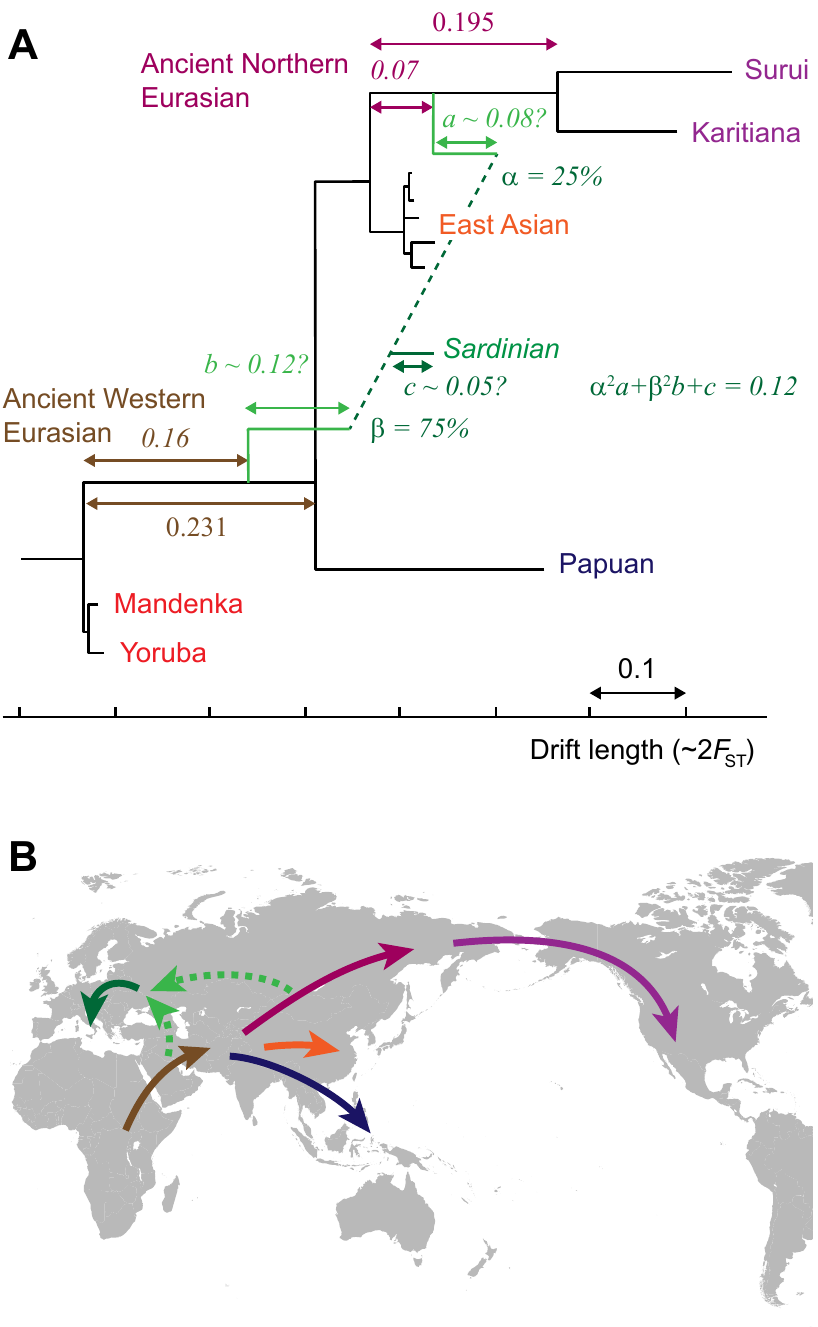}
\end{center}
\caption{{\bf Inferred ancient admixture in Europe.}  (A) Detail of the inferred ancestral admixture for Sardinians (other European populations are similar).  One mixing population splits from the unadmixed tree along the common ancestral branch of Native Americans (``Ancient Northern Eurasian'') and the other along the common ancestral branch of all non-Africans (``Ancient Western Eurasian'').  Median parameter values are shown; 95\% bootstrap confidence intervals can be found in Table~\ref{tab:european_mixes}.  The branch lengths $a$, $b$, and $c$ are confounded, so we show a plausible combination.  (B) Map showing a sketch of possible directions of movement of ancestral populations.  Colored arrows correspond to labeled branches in (A).}
\label{fig:european_detail}
\end{figure}

Using \MixMapper, we added European populations to the unadmixed scaffold via admixtures (Figure~\ref{fig:european_detail}; Table~\ref{tab:european_mixes}). 
For all eight groups in the HGDP data set, the best fit was as a mixture of a population related to the common ancestor of Karitiana and Suru\'{i} (in varying proportions of about 20--40\%, with Sardinian and Basque among the lowest and Russian the highest) with a population related to the common ancestor of all non-African populations on the tree.  We fit all eight European populations independently, but notably, their ancestors branch from the scaffold tree at very similar points, suggesting a similar broad-scale history.  Their branch positions are also qualitatively consistent with previous work that used the 3-population test to deduce ancient admixture for Europeans other than Sardinian and Basque~\citep{draft7}.  To confirm the signal in Sardinian and Basque, we applied $f_4$ ratio estimation~\citep{India,draft7}, which uses allele frequency statistics in a simpler framework to infer mixture proportions.  We estimated approximately 20--25\% ``ancient northern Eurasian'' ancestry (Table~\ref{tab:f4_ratio_Sardinian_Basque}), which is in very good agreement with our findings from \MixMapper (Table~\ref{tab:european_mixes}).

At first glance, this inferred admixture might appear improbable on geographical and chronological grounds, but importantly, the two ancestral branch positions do not represent the mixing populations themselves.  Rather, there may be substantial drift from the best-fit branch points to the true mixing populations, indicated as branch lengths $a$ and $b$ in Figure~\ref{fig:european_detail}A.  Unfortunately, these lengths, along with the post-admixture drift $c$, appear only in a fixed linear combination in the system of $f_2$ equations (Text~\ref{text:f-stats+admixture}), and current methods can only give estimates of this linear combination rather than the individual values~\citep{draft7}.  One plausible arrangement, however, is shown in Figure~\ref{fig:european_detail}A for the case of Sardinian. %

\subsection*{Two-way admixtures outside of Europe}

We also found several other populations that fit robustly onto the unadmixed tree using simple two-way admixtures (Table~\ref{tab:other_single_mixes}).
All of these can be identified as admixed using the 3-population or 4-population tests~\citep{draft7}, but with \MixMapper, we are able to provide the full set of best-fit parameter values to model them in an admixture tree.

First, we found that four populations from North-Central and Northeast Asia---Daur, Hezhen, Oroqen, and Yakut---are likely descended from admixtures between native North Asian populations and East Asian populations related to Japanese.  The first three are estimated to have roughly 10--30\% North Asian ancestry, while Yakut has 50--75\%.  Melanesians fit optimally as a mixture of a Papuan-related population with an East Asian population close to Dai, in a proportion of roughly 80\% Papuan-related, similar to previous estimates~\citep{reich2011denisova,xu2012genetic}. %
Finally, we found that Han Chinese have an optimal placement as an approximately equal mixture of two ancestral East Asian populations, one related to modern Dai (likely more southerly) and one related to modern Japanese (likely more northerly), corroborating a previous finding of admixture in Han populations between northern and southern clusters in a large-scale genetic analysis of East Asia~\citep{majumder2009mapping}.

\subsection*{Recent three-way admixtures involving western Eurasians} %

Finally, we inferred the branch positions of several populations that are well known to be recently admixed (cf.~\citet{draft7,pickrell2012inference}) but for which one ancestral mixing population was itself anciently admixed in a similar way to Europeans.  To do so, we applied the capability of \MixMapper to fit three-way admixtures (Figure~\ref{fig:table_key}B), using the anciently admixed branch leading to Sardinian as one ancestral source branch. First, we found that Mozabite, Bedouin, Palestinian, and Druze, in decreasing order of African ancestry, are all optimally represented as a mixture between an African population and an admixed western Eurasian population (not necessarily European) related to Sardinian (Table~\ref{tab:double_mixes}).  We also obtained good fits for Uygur and Hazara as mixtures between a western Eurasian population and a population related to the common ancestor of all East Asians on the tree (Table~\ref{tab:double_mixes}).

\subsection*{Estimation of ancestral heterozygosity}

\begin{figure}[H]
\begin{center}
\includegraphics{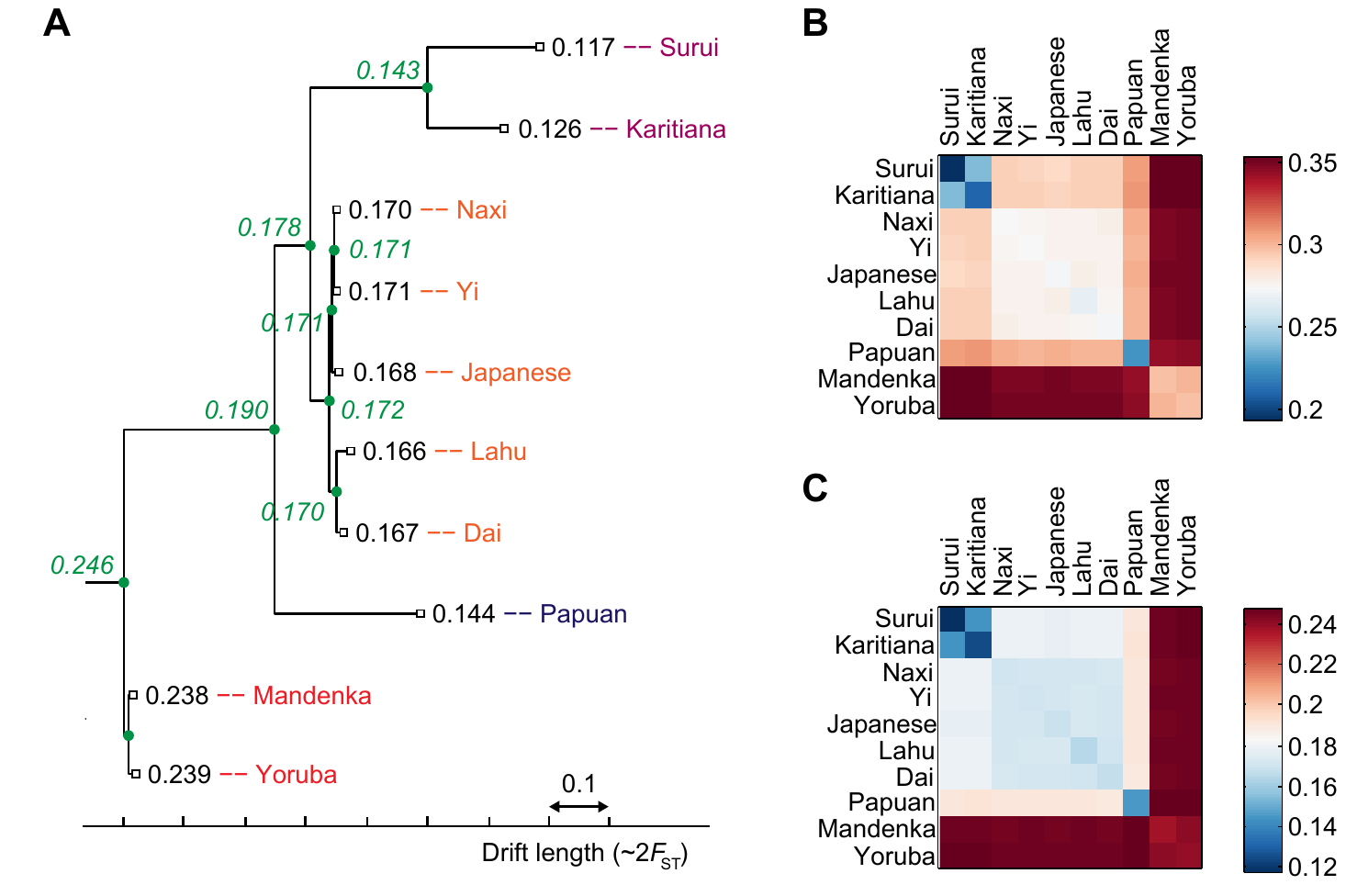}
\end{center}
\caption{{\bf Ancestral heterozygosity imputed from original Illumina vs.\ San-ascertained SNPs.}  (A) The 10-population unadmixed tree with estimated average heterozygosities using SNPs from Panel 4 (San ascertainment) of the Affymetrix Human Origins array~\citep{draft7}.  Numbers in black are direct calculations for modern populations, while numbers in green are inferred values at ancestral nodes.  (B, C) Computed ancestral heterozygosity at the common ancestor of each pair of modern populations.  With unbiased data, values should be equal for pairs having the same common ancestor. %
(B) Values from a filtered subset of about 250,000 SNPs from the published Illumina array data~\citep{li2008worldwide}. (C) Values from the Human Origins array excluding SNPs in gene regions.  }
\label{fig:het_tree_heatmaps}
\end{figure}

Using SNPs ascertained in an outgroup to all of our study populations enables us to compute accurate estimates of the heterozygosity (over a given set of SNPs) throughout an unadmixed tree, including at ancestral nodes (see Material and Methods).  This in turn allows us to convert branch lengths from $f_2$ units to easily interpretable drift lengths (see Text~\ref{text:het+drift}).
In Figure~\ref{fig:het_tree_heatmaps}C, we show our estimates for the heterozygosity (averaged over all San-ascertained SNPs used) at the most recent common ancestor (MRCA) of each pair of present-day populations in the tree.  Consensus values are given at the nodes of Figure~\ref{fig:het_tree_heatmaps}A.  The imputed heterozygosity should be the same for each pair of populations with the same MRCA, and indeed, with the new data set, the agreement is excellent (Figure~\ref{fig:het_tree_heatmaps}C).  %
By contrast, inferences of ancestral heterozygosity are much less accurate using HGDP data from the original Illumina SNP array~\citep{li2008worldwide} because of ascertainment bias (Figure~\ref{fig:het_tree_heatmaps}B); $f_2$ statistics are also affected but to a lesser degree (Figure~\ref{fig:f2_diff_heatmap}), as previously demonstrated~\citep{draft7}. We used these heterozygosity estimates to express branch lengths of all of our trees in drift units (Text~\ref{text:het+drift}).

\section*{Discussion}

\subsection*{Comparison with previous approaches}

The \MixMapper framework generalizes and automates several previous
admixture inference tools based on allele frequency moment statistics, incorporating them as
special cases. %
Methods such as the
3-population test for admixture and $f_4$ ratio estimation
\citep{India, draft7} have similar theoretical underpinnings, %
but \MixMapper provides more extensive information by analyzing more
populations simultaneously and automatically considering different
tree topologies and sources of gene flow. %
For example, negative $f_3$ values---i.e., 3-population tests
indicating admixture---can be expressed in terms of relationships
among $f_2$ distances between populations in an admixture tree.  In
general, 3-population tests can be somewhat difficult to interpret
because the surrogate ancestral populations may not in fact be closely
related to the true participants in the admixture, e.g., in the
``outgroup case'' \citep{India, draft7}.  The relations among the
$f_2$ statistics incorporate this situation naturally, however, and
solving the full system recovers the true branch points wherever they
are.  As another example, $f_4$ ratio estimation infers mixture
proportions of a single admixture event from $f_4$ statistics
involving the admixed population and four unadmixed populations
situated in a particular topology~\citep{India, draft7}.  Whenever
data for five such populations are available, the system of all $f_2$
equations that \MixMapper solves to obtain the mixture fraction
becomes equivalent to the $f_4$ ratio computation.  More importantly,
because \MixMapper infers all of the topological relationships within
an admixture tree automatically by optimizing the solution of the
distance equations over all branches, we do not need to specify in
advance where the admixture took place---which is not always obvious
{\em a priori}.  By using more than five populations, \MixMapper also
benefits from more data points to constrain the fit.

\MixMapper also offers significant advantages over the \emph{qpgraph} admixture tree fitting software \citep{draft7}.  Most notably, \emph{qpgraph} requires the user to specify the entire topology of the tree, including admixtures, in advance.  This requires either prior knowledge of sources of gene flow relative to the reference populations or a potentially lengthy search to test alternative branch locations.  \MixMapper is also faster and provides the capabilities to convert branch lengths into drift units and to perform bootstrap replicates to measure uncertainty in parameter estimates.  Furthermore, \MixMapper is designed to have more flexible and intuitive input and output and better diagnostics for incorrectly specified models.  While \emph{qpgraph} does fill a niche of fitting very precise models for small sets of populations, it becomes quite cumbersome for more than about seven or eight, whereas \MixMapper can be run with significantly larger trees without sacrificing efficiency, ease of use, or accuracy of inferences for populations of interest.

Finally, \MixMapper differs from \TreeMix~\citep{pickrell2012inference} in its emphasis on precise and flexible modeling of individual admixed populations.  
Stylistically, we view \MixMapper as ``semi-automated'' as compared to \TreeMix, which
is almost fully automated.  Both approaches have benefits: ours allows
more manual guidance and lends itself to interactive use, whereas
\TreeMix requires less user intervention, although some care must be
taken in choosing the number of gene flow events to include (10 in the HGDP results shown in
Figure~\ref{fig:TreeMix}) to avoid creating spurious mixtures.
With \MixMapper, we create admixture
trees including pre-selected approximately unadmixed populations together with admixed populations of interest, which are added on a case-by-case
basis only if they fit reliably as two- or three-way admixtures.  In
contrast, \TreeMix returns a single large-scale admixture tree
containing all populations in the input data set, which may include some
that can be shown to be admixed by other means but are not modeled as such.  
Thus, these populations might not be placed well on the tree, which in turn could affect the accuracy of the inferred admixture events.  Likewise, the populations ultimately modeled as admixed are initially included as part of an unadmixed tree, where (presumably) they do not fit well, which could introduce errors in the starting tree topology that impact the final results. 

\begin{figure}[H]
\begin{center}
\includegraphics{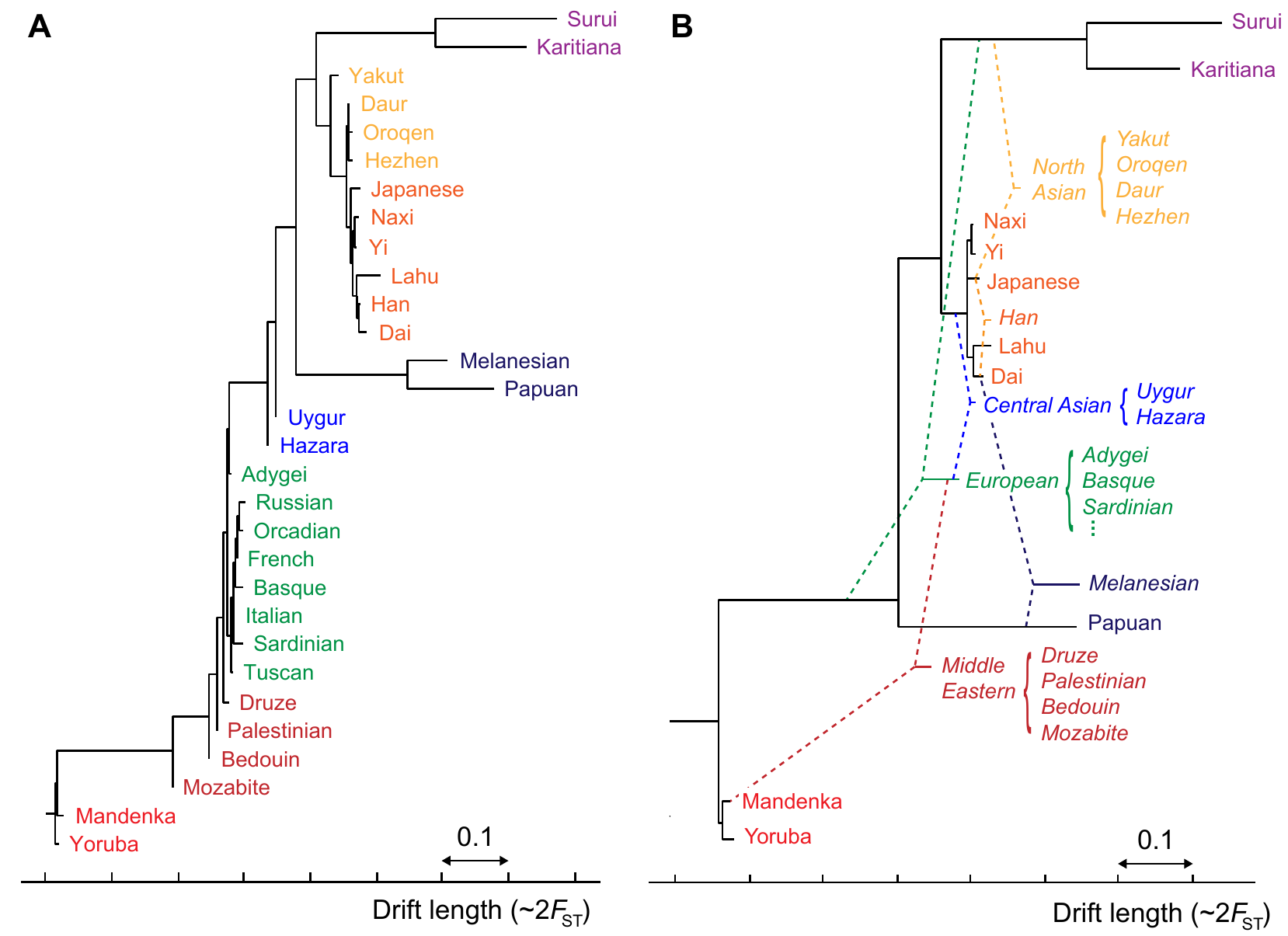}
\end{center}
\caption{{\bf Aggregate phylogenetic trees of HGDP populations with and without admixture.} (A) A simple neighbor-joining tree on the 30 populations for which \MixMapper produced high-confidence results.  This tree is analogous to the one given by \citet[Figure~1B]{li2008worldwide}, and the topology is very similar. (B) Results from \MixMapper.  The populations appear in roughly the same order, but the majority are inferred to be admixed, as represented by dashed lines (cf.~\citet{pickrell2012inference} and Figure~\ref{fig:TreeMix}). Note that drift units are not additive, so branch lengths should be interpreted individually.}%
\label{fig:mix_tree}
\end{figure}

Indeed, these methodological differences can be seen to affect inferences for both simulated and real data.  For our second simulated admixture tree, \MixMapper very accurately fit the populations with complicated histories (meant to mimic European and Middle Eastern populations), whereas \TreeMix only recovered portions of the true tree and also added two inaccurate mixtures (Figure~\ref{fig:simulations}).  We believe \TreeMix was hindered in this case by attempting to fit all of the populations simultaneously and by starting with all of them in an unadmixed tree.  In particular, once pop9 (with the lowest proportion of pop7-related admixture) was placed on the unadmixed tree, it likely became difficult to detect as admixed, while pop8's initial placement higher up the tree was likely due to its pop2-related admixture but then obscured this signal in the mixture-fitting phase.  Finally, the initial tree shape made populations 3-10 appear to be unequally drifted.
Meanwhile, with the HGDP data (Figures~\ref{fig:mix_tree} and~\ref{fig:TreeMix}), both methods fit
Palestinian, Bedouin, Druze, Mozabite, Uygur, and Hazara as
admixed, but \MixMapper analysis suggested that these populations
are better modeled as three-way admixed.  \TreeMix alone fit Brahui,
Makrani, Cambodian, and Maya---all of which the
3-population test identifies as admixed but we were unable to place
reliably with \MixMapper---while \MixMapper alone confidently fit Daur, Hezhen, Oroqen, Yakut, Melanesian, and Han.
Perhaps most notably, \MixMapper alone inferred
widespread ancient admixture for Europeans; the closest possible signal of such an event in the \TreeMix model is a migration edge from an ancestor of Native Americans to Russians.
We believe that, as in the simulations, \MixMapper is better suited to finding a common, ancient admixture signal in a group of populations, and more generally to disentangling complex admixture signals from within a large set of populations, and hence it is able to detect admixture in Europeans when \TreeMix does not.

To summarize, \MixMapper offers a suite of features that make it better suited than existing methods for the purpose of inferring accurate admixture parameters in data sets containing many specific populations of interest.  Our approach provides a middle ground between \emph{qpgraph}, which is designed to fit small numbers of populations within almost no residual errors, and \TreeMix, which generates large trees with little manual intervention but may be less precise in complex admixture scenarios.  Moreover, \MixMapper's %
speed and interactive design allow the user to evaluate the uncertainty and robustness of results in ways that we have found to be very useful (e.g., by comparing two- vs.\ three-way admixture models or results obtained using alternative scaffold trees).

\subsection*{Ancient European admixture}

Due in part to the flexibility of the \MixMapper approach, we were able to obtain the notable result that all European populations in the HGDP are best modeled as mixtures between a population related to the common ancestor of Native Americans and a population related to the common ancestor of all non-African populations in our scaffold tree, confirming and extending an admixture signal first reported by~\citet{draft7}. Our interpretation is that most if not all modern Europeans are descended from at least one large-scale ancient admixture event involving, in some combination, at least one population of Mesolithic European hunter-gatherers; Neolithic farmers, originally from the Near East; and/or other migrants from northern or Central Asia. Either the first or second of these could be related to the ``ancient western Eurasian'' branch in Figure~\ref{fig:european_detail}, and either the first or third could be related to the ``ancient northern Eurasian'' branch. %
Present-day Europeans differ in the amount of drift they have experienced since the admixture and in the proportions of the ancestry components they have inherited, but their overall profiles are similar.

Our results for Europeans are consistent with several previously
published lines of evidence~\citep{pinhasi2012genetic}.  First, it
has long been hypothesized, based on analysis of a few genetic loci
(especially on the Y chromosome), that Europeans are descended from
ancient
admixtures~\citep{semino2000genetic,dupanloup2004estimating,soares2010archaeogenetics}.
Our results also suggest an interpretation for a previously unexplained
\emph{frappe} analysis of worldwide human population structure (using
$K=4$ clusters) showing that almost all Europeans contain a small
fraction of American-related ancestry~\citep{li2008worldwide}.
Finally, sequencing of ancient DNA has revealed substantial
differentiation in Neolithic Europe between farmers and
hunter-gatherers~\citep{bramanti2009genetic}, with the former more
closely related to present-day Middle
Easterners~\citep{haak2010ancient} and southern
Europeans~\citep{keller2012new,skoglund2012origins} and the latter
more similar to northern Europeans~\citep{skoglund2012origins}, a
pattern perhaps reflected in our observed northwest-southeast cline in
the proportion of ``ancient northern Eurasian'' ancestry
(Table~\ref{tab:european_mixes}). Further analysis of ancient DNA may
help shed more light on the sources of ancestry of modern Europeans~\citep{der2013ancient}.

One important new insight of our European analysis is that we detect
the same signal of admixture in Sardinian and Basque as in the rest of
Europe.  As discussed above, unlike other Europeans, Sardinian and
Basque cannot be confirmed to be admixed using the 3-population test
(as in~\citet{draft7}), likely due to a combination of less ``ancient
northern Eurasian'' ancestry and more genetic drift since the
admixture (Table~\ref{tab:european_mixes}).  The first point is
further complicated by the fact that we have no unadmixed ``ancient
western Eurasian'' population available to use as a reference; indeed,
Sardinians themselves are often taken to be such a reference.
However, \MixMapper uncovered strong evidence for admixture in
Sardinian and Basque through additivity-checking in the first phase of the program
and automatic topology optimization in the second phase, discovering
the correct arrangement of unadmixed populations and
enabling admixture parameter inference, which we then verified
directly with $f_4$ ratio estimation.  Perhaps the most convincing
evidence of the robustness of this finding is that
\MixMapper infers branch points for the ancestral mixing populations
that are very similar to those of other Europeans
(Table~\ref{tab:european_mixes}), a concordance that is most
parsimoniously explained by a shared history of ancient admixture
among Sardinian, Basque, and other European populations.  Finally, we
note that because we fit all European populations without assuming Sardinian or Basque to be an unadmixed
reference, our estimates of the ``ancient northern Eurasian'' ancestry
proportions in Europeans are larger than those in~\citet{draft7} and
we believe more accurate than others previously reported~\citep{skoglund2012origins}.

\subsection*{Future directions}

It is worth noting that of the 52 populations (excluding San) in the HGDP data set, there were 22 that we were unable to fit in a reasonable way either on the unadmixed tree or as admixtures.  %
In part, this was because our instantaneous-admixture model is intrinsically limited in its ability to capture complicated population histories.  Most areas of the world have surely witnessed ongoing low levels of inter-population migration over time, especially between nearby populations, making it difficult to fit admixture trees to the data.   %
We also found cases where having data from more populations would help the fitting process, for example for three-way admixed populations such as Maya where we do not have a sampled group with a simpler admixture history that could be used to represent two of the three components.
Similarly, we found that while Central Asian populations such as Burusho, Pathan, and Sindhi have clear signals of admixture from the 3-population test, their ancestry can likely be traced to several different sources (including sub-Saharan Africa in some instances), making them difficult to fit with \MixMapper, particularly using the HGDP data.  Finally, we have chosen here to disregard admixture with archaic humans, which is known to be a small but noticeable component for most populations in the HGDP~\citep{green2010draft, reich2010genetic}.  In the future, it will be interesting to extend \MixMapper and other admixture tree-fitting methods to incorporate the possibilities of multiple-wave and continuous admixture. %

In certain applications, full genome sequences are beginning to replace more limited genotype data sets such as ours, but we believe that our methods and SNP-based inference in general will still be valuable in the future.  Despite the improving cost-effectiveness of sequencing, it is still much easier and less expensive to genotype samples using a SNP array, and with over 100,000 loci, the data used in this study provide substantial statistical power.  Additionally, sequencing technology is currently more error-prone, which can lead to biases in allele frequency-based statistics~\citep{pool2010population}.%
We expect that \MixMapper will continue to contribute to an important toolkit of population history inference methods based on SNP allele frequency data.

\section*{Material and Methods}

\subsection*{Model assumptions and $f$-statistics}

We assume that all SNPs are neutral,
biallelic, and autosomal, and that divergence times are short enough
that there are no double mutations at a locus. Thus, allele frequency
variation---the signal that we harness---is governed entirely by
genetic drift and admixture.
We model admixture as a one-time exchange of genetic material: two
parent populations mix to form a single descendant population whose
allele frequencies are a weighted average of the parents'.  This model
is of course an oversimplification of true mixture events, but it is
flexible enough to serve as a first-order approximation.

Our point-admixture model is amenable to allele frequency moment analyses based on $f$-statistics~\citep{India,draft7}. We primarily make use of the statistic $f_2(A,B) := E_S[(p_A-p_B)^2]$, where $p_A$ and $p_B$ are allele frequencies in populations $A$ and $B$, and $E_S$ denotes the mean over all SNPs.  Expected values of $f_2$ can be written in terms of admixture tree parameters as described in Text~\ref{text:f-stats+admixture}.  Linear combinations of $f_2$ statistics can also be used to form the quantities $f_3(C;A,B) := E_S[(p_C-p_A)(p_C-p_B)]$ and $f_4(A,B;C,D) := E_S[(p_A-p_B)(p_C-p_D)]$, which form the bases of the 3- and 4-population tests for admixture, respectively.  For all of our $f$-statistic computations, we use previously described unbiased estimators~\citep{India,draft7}.

\subsection*{Constructing an unadmixed scaffold tree}

Our \MixMapper admixture-tree-building procedure consists of two
phases (Figure~\ref{fig:flow_chart}), the first of which selects a set
of unadmixed populations to use as a scaffold tree.  We
begin by computing $f_3$ statistics~\citep{India, draft7} for all
triples of populations $P_1, P_2, P_3$ in the data set and removing
those populations $P_3$ with any negative values $f_3(P_3; P_1, P_2)$,
which indicate admixture.  We then use pairwise $f_2$ statistics to
build neighbor-joining trees on subsets of the remaining
populations.  In the absence of admixture, $f_2$ distances are
additive along paths on a phylogenetic tree (Text~\ref{text:f-stats+admixture};
cf.~\citet{draft7}), meaning that neighbor-joining should recover a
tree with leaf-to-leaf distances that are completely consistent with
the pairwise $f_2$ data~\citep{saitou1987neighbor}.  However, with real data, the putative unadmixed subsets are rarely completely additive, meaning that the fitted neighbor-joining trees have residual errors between the inferred leaf-to-leaf distances and the true $f_2$ statistics.  These deviations from additivity are equivalent to non-zero results from the 4-population test for admixture~\citep{India,draft7}.  We therefore
evaluate the quality of each putative unadmixed tree according to its
maximum error between fitted and actual pairwise distances: for a tree $\mathit{T}$ having distances $d$ between populations $P$ and $Q$, the deviation from additivity is defined as $\max\{|d(P,Q)-f_2(P,Q)|:P,Q \in \mathit{T}\}$.  
\MixMapper computes this deviation on putatively unadmixed subsets of increasing size, retaining a user-specified number of best subsets of each size in a ``beam search" procedure to avoid exponential complexity.

Because of model violations in real data, trees built on smaller subsets are
more additive, but they are also less informative; in particular, it
is beneficial to include populations from as many continental groups
as possible in order to provide more potential branch points for admixture
fitting.  \MixMapper provides a ranking of the most additive trees of each size as a
guide from which the user chooses a suitable unadmixed scaffold.  
Once the rank-list of trees has been generated, subject to some constraints (e.g., certain populations required), the user can scan the first several most additive trees for a range of sizes, looking for a balance between coverage and accuracy.  This can also be accomplished by checking whether removing a population from a proposed tree results in a substantial additivity benefit; if so, it may be wise to eliminate it.  Similarly, if the population removed from the tree can be modeled well as admixed using the remaining portion of the scaffold, this provides evidence that it should not be part of the unadmixed tree. Finally,
\MixMapper adjusts the scaffold tree that the user ultimately selects by re-optimizing its branch lengths
(maintaining the topology inferred from neighbor-joining) to minimize
the sum of squared errors of all pairwise $f_2$ distances.

Within the above guidelines, users should choose the scaffold tree most appropriate for their purposes, which may involve other considerations.  In addition to additivity and overall size, it is sometimes desirable to select more or fewer populations from certain geographical, linguistic, or other categories.  For example, including a population in the scaffold that is actually admixed might not affect the inferences as long as it is not too closely related to the admixed populations being modeled.  At the same time, it can be useful to have more populations in the scaffold around the split points for an admixed population of interest in order to obtain finer resolution on the branch positions of the mixing populations.  For human data in particular, the unadmixed scaffold is only a modeling device; the populations it contains likely have experienced at least a small amount of mixture.  A central goal in building the scaffold is to choose populations such that applying this model will not interfere with the conclusions obtained using the program.  The interactive design of \MixMapper allows the user to tweak the scaffold tree very easily in order to check robustness, and in our analyses, conclusions are qualitatively unchanged for different scaffolds (Figures~\ref{fig:alt_scaffold}--\ref{fig:alt_alphas}; Tables~\ref{tab:european_mixes_alt1}--\ref{tab:double_mixes_alt1}).

\subsection*{Two-way admixture fitting}

\begin{figure}[H]
\begin{center}
\includegraphics{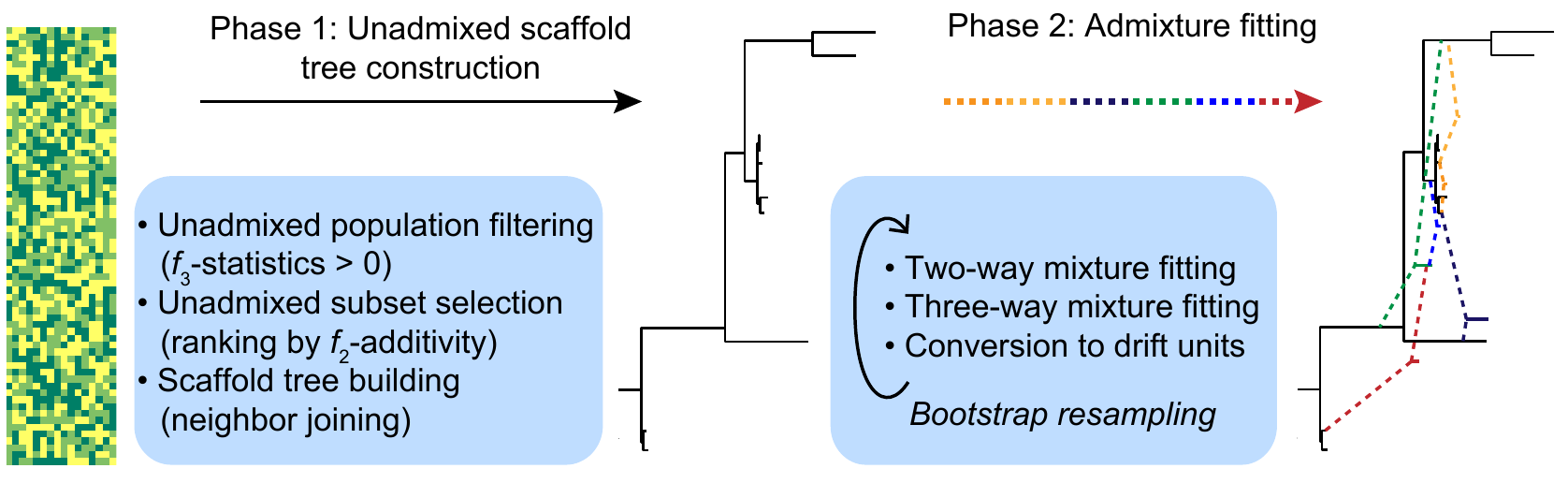}
\end{center}
\caption{{\bf \MixMapper workflow.}  \MixMapper takes as input an array of SNP calls annotated with the population to which each individual belongs.  The method then proceeds in two phases, first building a tree of (approximately) unadmixed populations and then attempting to fit the remaining populations as admixtures.  In the first phase, \MixMapper produces a ranking of possible unadmixed trees in order of deviation from $f_2$-additivity; based on this list, the user selects a tree to use as a scaffold.  In the second phase, \MixMapper tries to fit remaining populations as two- or three-way mixtures between branches of the unadmixed tree.  In each case \MixMapper produces an ensemble of predictions via bootstrap resampling, enabling confidence estimation for inferred results.}

\label{fig:flow_chart}
\end{figure}

The second phase of \MixMapper begins by attempting to fit
additional populations independently as simple two-way admixtures
between branches of the unadmixed tree (Figure~\ref{fig:flow_chart}).
For a given admixed population, assuming for the moment that we know the branches from which the
ancestral mixing populations split, we can construct a system of
equations of $f_2$ statistics that allows us to infer parameters of
the mixture (Text~\ref{text:f-stats+admixture}).  Specifically, the
squared allele frequency divergence $f_2(M,X')$ between the admixed population $M$ and
each unadmixed population $X'$ can be expressed as an algebraic
combination of known branch lengths along with four unknown mixture
parameters: the locations of the split points on the two parental
branches, the combined terminal branch length, and the mixture
fraction (Figure~\ref{fig:table_key}A).  To solve for the four
unknowns, we need at least four unadmixed populations $X'$ that
produce a system of four independent constraints on the parameters.
This condition is satisfied if and only if the data set contains two
populations $X'_1$ and $X'_2$ that branch from different points along
the lineage connecting the divergence points of the parent populations
from the unadmixed tree (Text~\ref{text:f-stats+admixture}).  If the
unadmixed tree contains $n > 4$ populations, we obtain a system of $n$
equations in the four unknowns that in theory is dependent.  In
practice, the equations are in fact slightly inconsistent because of
noise in the $f_2$ statistics and error in the point-admixture model,
so we perform least-squares optimization to solve for the unknowns;
having more populations helps reduce the impact of noise.

Algorithmically, \MixMapper performs two-way admixture fitting by
iteratively testing each pair of branches of the unadmixed tree as
possible sources of the two ancestral mixing populations.  For each choice of branches,
\MixMapper builds the implied system of equations and finds the
least-squares solution (under the constraints that unknown branch
lengths are nonnegative and the mixture fraction $\alpha$ is between 0
and 1), ultimately choosing the pair of branches and mixture
parameters producing the smallest residual norm.  Our procedure for
optimizing each system of equations uses the observation that upon
fixing $\alpha$, the system becomes linear in the
remaining three variables (Text~\ref{text:f-stats+admixture}).  Thus,
we can optimize the system by performing constrained linear least
squares within a basic one-parameter optimization routine over $\alpha
\in [0, 1]$.  To implement this approach, we applied MATLAB's
\texttt{lsqlin} and \texttt{fminbnd} functions with a few auxiliary
tricks to improve computational efficiency (detailed in the code).

\subsection*{Three-way admixture fitting}

\MixMapper also fits three-way admixtures, i.e., those for which one
parent population is itself admixed (Figure~\ref{fig:table_key}B).
Explicitly, after an admixed population $M_1$ has been added to the
tree, \MixMapper can fit an additional user-specified admixed
population $M_2$ as a mixture between the $M_1$ terminal branch and
another (unknown) branch of the unadmixed tree.  The fitting algorithm
proceeds in a manner analogous to the two-way mixture case: \MixMapper
iterates through each possible choice of the third branch, optimizing
each implied system of equations expressing $f_2$ distances in terms
of mixture parameters.  With two admixed populations, there are now
$2n+1$ equations, relating observed values of $f_2(M_1, X')$ and $f_2(M_2, X')$ for all unadmixed
populations $X'$, and also $f_2(M_1, M_2)$, to eight unknowns: two
mixture fractions, $\alpha_1$ and $\alpha_2$, and six branch
length parameters (Figure~\ref{fig:table_key}B).  Fixing $\alpha_1$
and $\alpha_2$ results in a linear system as before, so we perform the
optimization using MATLAB's \texttt{lsqlin} within \texttt{fminsearch}
applied to $\alpha_1$ and $\alpha_2$ in tandem.  The same mathematical
framework could be extended to optimizing the placement of populations with
arbitrarily many ancestral admixture events, but for simplicity and
to reduce the risk of overfitting, we chose to limit this version of
\MixMapper to three-way admixtures.

\subsection*{Expressing branch lengths in drift units}

All of the tree-fitting computations described thus far are performed using pairwise distances in $f_2$ units, which are mathematically convenient to work with owing to their additivity along a lineage (in the absence of admixture).  However, $f_2$ distances are not directly interpretable in the same way as genetic drift $D$, which is a simple function of time and population size:
\[
D \approx 1 - \exp(-t/2N_e) \approx 2 \cdot F_{ST},
\]
where $t$ is the number of generations and $N_e$ is the effective
population size~\citep{nei1987molecular}.
To convert $f_2$ distances to
drift units, we apply a new formula, dividing twice the $f_2$-length of each branch
by the heterozygosity value that we infer for the ancestral population
at the top of the branch (Text~\ref{text:het+drift}).  Qualitatively speaking, this conversion corrects for the relative stretching of $f_2$ branches at different portions of the tree as a function of heterozygosity~\citep{draft7}.  In order to infer ancestral heterozygosity values accurately, it is critical to use SNPs that are ascertained in an outgroup to the populations involved, which we address further below.

Before inferring heterozygosities at ancestral nodes of the unadmixed
tree, we must first determine the location of the root (which is
neither specified by neighbor-joining nor involved in the preceding
analyses).  \MixMapper does so by iterating through branches of the
unadmixed tree, temporarily rooting the tree along each branch, and
then checking for consistency of the resulting heterozygosity
estimates.  Explicitly, for each internal node $P$, we split its
present-day descendants (according to the re-rooted tree) into two
groups $G_1$ and $G_2$ according to which child branch of $P$ they
descend from.  For each pair of descendants, one from $G_1$ and one
from $G_2$, we compute an inferred heterozygosity at $P$
(Text~\ref{text:het+drift}).  If the tree is rooted properly, these
inferred heterozygosities are consistent, but if not, there exist
nodes $P$ for which the heterozygosity estimates conflict.  \MixMapper thus infers the location of the root as well as the
ancestral heterozygosity at each internal node, after which it applies
the drift length conversion as a post-processing step on fitted $f_2$
branch lengths.

\subsection*{Bootstrapping}

In order to measure the statistical significance of our parameter estimates, we compute bootstrap confidence intervals~\citep{efron1979bootstrap, efron1986bootstrap} for the inferred branch lengths and mixture fractions.  Our bootstrap procedure is designed to account for both the randomness of the drift process at each SNP and the random choice of individuals sampled to represent each population.  First, we divide the genome into 50 evenly-sized blocks, with the premise that this scale should easily be larger than that of linkage disequilibrium among our SNPs.  Then, for each of 500 replicates, we resample the data set by (a) selecting 50 of these SNP blocks at random with replacement; and (b) for each population group, selecting a random set of individuals with replacement, preserving the number of individuals in the group.

For each replicate, we recalculate all pairwise $f_2$ distances and present-day heterozygosity values using the resampled SNPs and individuals (adjusting the bias-correction terms to account for the repetition of individuals) and then construct the admixture tree of interest. %
Even though the mixture parameters we estimate---branch lengths and mixture fractions---depend in complicated ways on many different random variables, we can directly apply the nonparametric bootstrap to obtain confidence intervals~\citep{efron1986bootstrap}.  For simplicity, we use a percentile bootstrap; thus, our 95\% confidence intervals indicate 2.5 and 97.5 percentiles of the distribution of each parameter among the replicates.

Computationally, we parallelize \MixMapper's mixture-fitting over the bootstrap replicates using MATLAB's Parallel Computing Toolbox.

\subsection*{Evaluating fit quality}

When interpreting admixture inferences produced by methods such as \MixMapper, it is important to ensure that best-fit models are in fact accurate.  While formal tests for goodness of fit do not generally exist for methods of this class, we
use several criteria to evaluate the mixture fits produced by \MixMapper and distinguish high-confidence results from possible artifacts of overfitting or model violations.  

First, we can compare \MixMapper results to information obtained from other methods, such as the 3-population test~\citep{India, draft7}. Negative $f_3$ values indicate robustly that the tested population is admixed, and comparing $f_3$ statistics for different reference pairs can give useful clues about the ancestral mixing populations.  Thus, while the 3-population test relies on similar data to \MixMapper, its simpler form makes it useful for confirming that \MixMapper results are reasonable.   %

Second, the consistency of parameter values over bootstrap replicates gives an indication of the robustness of the admixture fit in question.  All results with real data have some amount of associated uncertainty, which is a function of sample sizes, SNP density, intra-population homogeneity, and other aspects of the data.  Given these factors, we place less faith in results with unexpectedly large error bars.  Most often, this phenomenon is manifested in the placement of ancestral mixing populations: for poorly fitting admixtures, branch choices often change from one replicate to the next, signaling unreliable results.%

Third, we find that results where one ancestral population is very closely related to the admixed population and contributes more than 90\% of the ancestry are often unreliable.   We expect that if we try to fit a non-admixed population as an admixture, \MixMapper should return a closely related population as the first branch with mixture fraction $\alpha \approx 1$ (and an arbitrary second branch).  Indeed, we often observe this pattern in the context of verifying that certain populations make sense to include in the scaffold tree.
Further evidence of overfitting comes when the second ancestry component, which contributes only a few percent, either bounces from branch to branch over the replicates, is located at the very tip of a leaf branch, or is historically implausible.   

Fourth, for any inferred admixture event, the two mixing populations must be contemporaneous.  Since we cannot resolve the three pieces of terminal drift lengths leading to admixed populations (Figure~\ref{fig:table_key}A) and our branch lengths depend both on population size and absolute time, we cannot say for sure whether this property is satisfied for any given mixture fit.  In some cases, however, it is clear that no realization of the variables could possibly be consistent: for example, if we infer an admixture between a very recent branch and a very old one with a small value of the total mixed drift---and hence the terminal drift $c$---then we can confidently say the mixture is unreasonable.

Finally, when available, we also use prior historical or other external knowledge to guide what we consider to be reasonable.  Sometimes, the model that appears to fit the data best has implications that are clearly historically implausible; often when this is true one or more of the evaluation criteria listed above can be invoked as well.  Of course, the most interesting findings are often those that are new and surprising, but we subject such results to an extra degree of scrutiny.

\subsection*{Data set and ascertainment}

We analyzed a SNP data set from 934 HGDP individuals grouped in 53 populations~\citep{rosenberg2002genetic, li2008worldwide}.  Unlike most previous studies of the HGDP samples, however, we worked with recently published data generated using the new Affymetrix Axiom Human Origins Array~\citep{draft7}, which was designed with a simple ascertainment scheme for accurate population genetic inference~\citep{keinan2007measurement}.  It is well known that ascertainment bias can cause errors in estimated divergences among populations~\citep{clark2005ascertainment, albrechtsen2010ascertainment}, since choosing SNPs based on their properties in modern populations induces non-neutral spectra in related samples.  While there do exist methods to correct for ascertainment bias~\citep{nielsen2004reconstituting}, it is much more desirable to work with \textit{a priori} bias-free data, especially given that typical SNP arrays are designed using opaque ascertainment schemes.

To avoid these pitfalls, we used Panel 4 of the new array, which consists of 163,313 SNPs that were ascertained as heterozygous in the genome of a San individual~\citep{keinan2007measurement}.  This panel is special because there is evidence that the San are approximately an outgroup to all other modern-day human populations~\citep{li2008worldwide, gronau2011bayesian}.  Thus, while the Panel 4 ascertainment scheme distorts the San allele frequency spectrum, it is nearly neutral with respect to all other populations.  In other words, we can think of the ascertainment as effectively choosing a set of SNPs (biased toward San heterozygosity) at the common ancestor of the remaining 52 populations, after which drift occurs in a bias-free manner.  We excluded 61,369 SNPs that are annotated as falling between the transcription start site and end site of a gene in the UCSC Genome Browser database~\citep{fujita2011ucsc}.  Most of the excluded SNPs are not within actual exons, but as expected, the frequency spectra at these ``gene region'' loci were slightly shifted toward fixed classes relative to other SNPs, indicative of the action of selection (Figure~\ref{fig:spectral_shift_near_genes}).  Since we assume neutrality in all of our analyses, we chose to remove these SNPs.

\subsection*{Simulations}

Our first simulated tree was generated using the \texttt{ms}~\citep{hudson2002generating} command
\begin{quote}
\footnotesize
\texttt{ms 350 500 -t 50 -r 99.9998 500000 -I 7 50 50 50 50 50 50 50 -n 7 2 -n 1 2 -n 2 2 -ej 0.04 2 1 -es 0.02 6 0.4 -ej 0.06 6 3 -ej 0.04 8 5 -ej 0.08 5 4 -ej 0.12 4 3 -ej 0.2 3 1 -ej 0.3 1 7 -en 0.3 7 1.}
\end{quote}
After ascertainment, we used a total of 95,997 SNPs.

Our second simulated tree was generated with the command
\begin{quote}
\footnotesize
\texttt{ms 550 500 -t 50 -r 99.9998 500000 -I 11 50 50 50 50 50 50 50 50 50 50 50 -n 11 2 -n 1 2 -n 2 2 -em 0.002 4 3 253.8 -em 0.004 4 3 0 -es 0.002 8 0.2 -en 0.002 8 2 -ej 0.02 8 2 -ej 0.02 4 5 -ej 0.04 2 1 -ej 0.04 5 3 -es 0.04 12 0.4 -es 0.04 9 0.2 -em 0.042 10 9 253.8 -em 0.044 10 9 0 -ej 0.06 12 7 -ej 0.06 9 7 -ej 0.06 14 10 -ej 0.06 13 10 -ej 0.08 7 6 -ej 0.12 6 3 -ej 0.16 10 3 -ej 0.2 3 1 -ej 0.3 1 11 -en 0.3 11 1.}
\end{quote}
After ascertainment, we used a total of 96,258 SNPs.  When analyzing this data set in \TreeMix, we chose to fit a total of five admixtures based on the residuals of the pairwise distances (maximum of approximately 3 standard errors) and our knowledge that this is the number in the true admixture tree (in order to make for a fair comparison).

\subsection*{Software}

Source code for the \MixMapper software is available at \url{http://groups.csail.mit.edu/cb/mixmapper/}. %

\section*{Acknowledgments}
We would like to thank George Tucker and Joe Pickrell for helpful discussions and the reviewers and editors for numerous suggestions that improved the manuscript.  %
This work was supported by the National Science Foundation (Graduate Research Fellowship support to M.L., P.L., and A.L.\ and HOMINID grant \#1032255 to D.R.\ and N.P.) and the National Institutes of Health (grant GM100233 to D.R.\ and N.P.)

\newpage
\bibliography{popgen}

\newpage
\section*{Figures}

\begin{figure}[H]
\begin{center}
\includegraphics{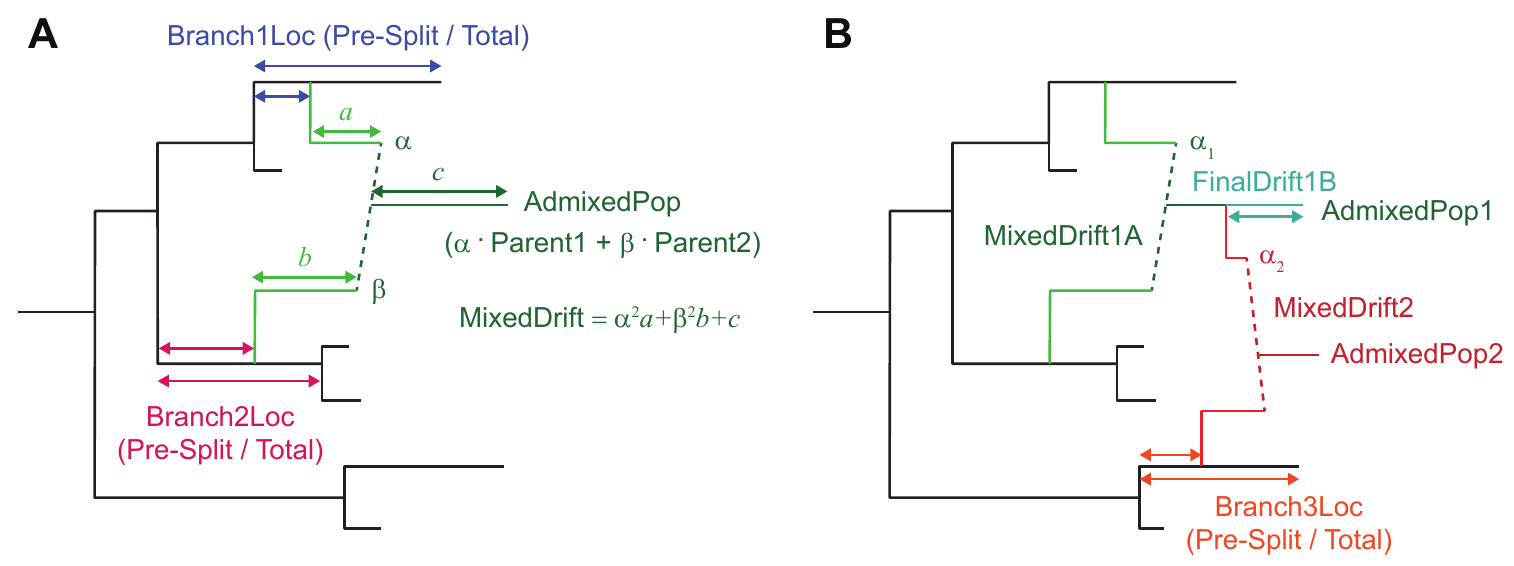}
\end{center}
\caption{{\bf Schematic of mixture parameters fit by \MixMapper.} (A) A simple two-way admixture.  \MixMapper infers four parameters when fitting a given population as an admixture.  It finds the optimal pair of branches between which to place the admixture and reports the following: Branch1Loc and Branch2Loc are the points at which the mixing populations split from these branches (given as pre-split length / total branch length); $\alpha$ is the proportion of ancestry from Branch1 ($\beta = 1-\alpha$ is the proportion from Branch2); and MixedDrift is the linear combination of drift lengths $\alpha^2a + \beta^2b + c$.  (B) A three-way mixture: here AdmixedPop2 is modeled as an admixture between AdmixedPop1 and Branch3.  There are now four additional parameters; three are analogous to the above, namely, Branch3Loc, $\alpha_2$, and MixedDrift2.  The remaining degree of freedom is the position of the split along the AdmixedPop1 branch, which divides MixedDrift into MixedDrift1A and FinalDrift1B.}
\label{fig:table_key}
\end{figure}

\newpage
\section*{Tables}

\begin{table}[H]
\caption{\bf{Mixture parameters for Europeans.}}
{\scriptsize\sffamily
\begin{tabular}{l r l l l l}
\hline
AdmixedPop & \# rep$^a$ & $\alpha$$^b$ & Branch1Loc (Anc.\ N.\ Eurasian)$^c$ & Branch2Loc (Anc.\ W.\ Eurasian)$^c$ & MixedDrift$^d$\\
\hline
Adygei           &  500 &  0.254-0.461  &  0.033-0.078 / 0.195  &  0.140-0.174 / 0.231  &  0.077-0.092             \\
Basque           &  464 &  0.160-0.385  &  0.053-0.143 / 0.196  &  0.149-0.180 / 0.231  &  0.105-0.121             \\
French           &  491 &  0.184-0.386  &  0.054-0.130 / 0.195  &  0.149-0.177 / 0.231  &  0.089-0.104             \\
Italian          &  497 &  0.210-0.415  &  0.043-0.108 / 0.195  &  0.137-0.173 / 0.231  &  0.092-0.109             \\
Orcadian         &  442 &  0.156-0.350  &  0.068-0.164 / 0.195  &  0.161-0.185 / 0.231  &  0.096-0.113             \\
Russian          &  500 &  0.278-0.486  &  0.045-0.091 / 0.195  &  0.146-0.181 / 0.231  &  0.079-0.095             \\
Sardinian        &  480 &  0.150-0.350  &  0.045-0.121 / 0.195  &  0.146-0.176 / 0.231  &  0.107-0.123             \\
Tuscan           &  489 &  0.179-0.431  &  0.039-0.118 / 0.195  &  0.137-0.177 / 0.231  &  0.088-0.110             \\
\hline
\end{tabular}
}
\begin{flushleft} %
$^a$Number of bootstrap replicates (out of 500) placing the mixture between the two branches shown.

$^b$Proportion of ancestry from ``ancient northern Eurasian'' (95\% bootstrap confidence interval).

$^c$See Figure~\ref{fig:european_detail}A for the definition of the ``ancient northern Eurasian'' and ``ancient western Eurasian'' branches in the scaffold tree; Branch1Loc and Branch2Loc are the points at which the mixing populations split from these branches (expressed as confidence interval for split point / branch total, as in Figure~\ref{fig:table_key}A).

$^d$Sum of drift lengths $\alpha^2a + (1-\alpha)^2b + c$; see Figure~\ref{fig:table_key}A.
\end{flushleft}
\label{tab:european_mixes}
\end{table}

\newpage
\begin{table}[H]
\caption{\bf{Mixture parameters for non-European populations modeled as two-way admixtures.}}
{\scriptsize\sffamily
\begin{tabular}{l l r l l l l}
\hline
AdmixedPop  & Branch1 + Branch2$^a$   &  \# rep$^b$ &  $\alpha$$^c$   &  Branch1Loc$^d$  &  Branch2Loc$^d$  &  MixedDrift$^e$ \\
\hline
Daur             & Anc.\ N.\ Eurasian + Japanese        &  350 &  0.067-0.276  &  0.008-0.126 / 0.195  &  0.006-0.013 / 0.016  &  0.006-0.015             \\
             & Suru\'{i} + Japanese               &  112 &  0.021-0.058  &  0.008-0.177 / 0.177  &  0.005-0.010 / 0.015  &  0.005-0.016             \\
\hline
Hezhen           & Anc.\ N.\ Eurasian + Japanese        &  411 &  0.068-0.273  &  0.006-0.113 / 0.195  &  0.006-0.013 / 0.016  &  0.005-0.029             \\
\hline
Oroqen           & Anc.\ N.\ Eurasian + Japanese        &  410 &  0.093-0.333  &  0.017-0.133 / 0.195  &  0.005-0.013 / 0.015  &  0.011-0.030             \\
           & Karitiana + Japanese           &   53 &  0.025-0.086  &  0.014-0.136 / 0.136  &  0.004-0.008 / 0.016  &  0.008-0.026             \\
\hline
Yakut            & Anc.\ N.\ Eurasian + Japanese        &  481 &  0.494-0.769  &  0.005-0.026 / 0.195  &  0.012-0.016 / 0.016  &  0.030-0.041             \\
\hline
Melanesian       & Dai + Papuan                   &  424 &  0.160-0.260  &  0.008-0.014 / 0.014  &  0.165-0.201 / 0.247  &  0.089-0.114             \\
       & Lahu + Papuan                  &   54 &  0.155-0.255  &  0.003-0.032 / 0.032  &  0.167-0.208 / 0.249  &  0.081-0.114             \\
\hline
Han              & Dai + Japanese                 &  440 &  0.349-0.690  &  0.004-0.014 / 0.014  &  0.008-0.016 / 0.016  &  0.002-0.006             \\
\hline
\end{tabular}
}
\begin{flushleft} %
$^a$Optimal split points for mixing populations.

$^b$Number of bootstrap replicates (out of 500) placing the mixture between Branch1 and Branch2; topologies are shown that that occur for at least 50 of 500 replicates.

$^c$Proportion of ancestry from Branch1 (95\% bootstrap confidence interval).

$^d$Points at which mixing populations split from their branches (expressed as confidence interval for split point / branch total, as in Figure~\ref{fig:table_key}A).

$^e$Sum of drift lengths $\alpha^2a + (1-\alpha)^2b + c$; see Figure~\ref{fig:table_key}A.
\end{flushleft}
\label{tab:other_single_mixes}
\end{table}

\newpage
\begin{table}[H]
\caption{\bf{Mixture parameters for populations modeled as three-way admixtures.}}
{\scriptsize\sffamily
\begin{tabular}{l l r l l l l l}
\hline
AdmixedPop2    & Branch3$^a$    &  \# rep$^b$ &  $\alpha_2$$^c$   &  Branch3Loc$^d$    &  MixedDrift1A$^e$ &  FinalDrift1B$^e$ &  MixedDrift2$^e$\\
\hline
Druze            & Mandenka           &  330 &  0.963-0.988  &  0.000-0.009 / 0.009  &  0.081-0.099  &  0.022-0.030  &  0.004-0.013     \\
            & Yoruba             &   82 &  0.965-0.991  &  0.000-0.010 / 0.010  &  0.080-0.099  &  0.022-0.029  &  0.005-0.013     \\
            & Anc.\ W.\ Eurasian          &   79 &  0.881-0.966  &  0.041-0.158 / 0.232  &  0.092-0.118  &  0.000-0.024  &  0.010-0.031     \\
\hline
Palestinian      & Anc.\ W.\ Eurasian          &  294 &  0.818-0.901  &  0.031-0.104 / 0.231  &  0.093-0.123  &  0.000-0.021  &  0.007-0.022     \\
      & Mandenka           &  146 &  0.909-0.937  &  0.000-0.009 / 0.009  &  0.083-0.097  &  0.022-0.029  &  0.001-0.007     \\
      & Yoruba             &   53 &  0.911-0.938  &  0.000-0.010 / 0.010  &  0.077-0.098  &  0.021-0.029  &  0.001-0.008     \\
\hline
Bedouin          & Anc.\ W.\ Eurasian          &  271 &  0.767-0.873  &  0.019-0.086 / 0.231  &  0.094-0.122  &  0.000-0.022  &  0.012-0.031     \\
          & Mandenka           &  176 &  0.856-0.923  &  0.000-0.008 / 0.008  &  0.080-0.099  &  0.023-0.030  &  0.006-0.018     \\
\hline
Mozabite         & Mandenka           &  254 &  0.686-0.775  &  0.000-0.009 / 0.009  &  0.088-0.109  &  0.012-0.022  &  0.017-0.032     \\
         & Anc.\ W.\ Eurasian          &  142 &  0.608-0.722  &  0.002-0.026 / 0.232  &  0.103-0.122  &  0.000-0.011  &  0.018-0.035     \\
         & Yoruba             &   73 &  0.669-0.767  &  0.000-0.008 / 0.010  &  0.086-0.108  &  0.012-0.023  &  0.017-0.031     \\
\hline
Hazara           & Anc.\ East Asian$^f$        &  497 &  0.364-0.471  &  0.010-0.024 / 0.034  &  0.080-0.115  &  0.004-0.034  &  0.004-0.013     \\
\hline
Uygur            & Anc.\ East Asian$^f$        &  500 &  0.318-0.438  &  0.007-0.023 / 0.034  &  0.088-0.123  &  0.000-0.027  &  0.000-0.009     \\
\hline
\end{tabular}
}
\begin{flushleft} %
$^a$Optimal split point for the third ancestry component.  The first two components are represented by a parent population splitting from the (admixed) Sardinian branch.

$^b$Number of bootstrap replicates placing the third ancestry component on Branch3; topologies are shown that that occur for at least 50 of 500 replicates.

$^c$Proportion of European-related ancestry (95\% bootstrap confidence interval).

$^d$Point at which mixing population splits from Branch3 (expressed as confidence interval for split point / branch total, as in Figure~\ref{fig:table_key}A).

$^e$Terminal drift parameters; see Figure~\ref{fig:table_key}B.

$^f$Common ancestral branch of the five East Asian populations in the unadmixed tree (Dai, Japanese, Lahu, Naxi, and Yi).
\end{flushleft}
\label{tab:double_mixes}
\end{table}

\newpage
\section*{Supplementary Materials: Efficient moment-based inference of admixture parameters and sources of gene flow}

Mark Lipson, Po-Ru Loh, Alex Levin, David Reich, Nick Patterson, and Bonnie Berger

\setcounter{figure}{0}
\renewcommand{\thefigure}{S\arabic{figure}}

\newpage
\begin{figure}[H]
\begin{center}
\includegraphics[width=\textwidth]{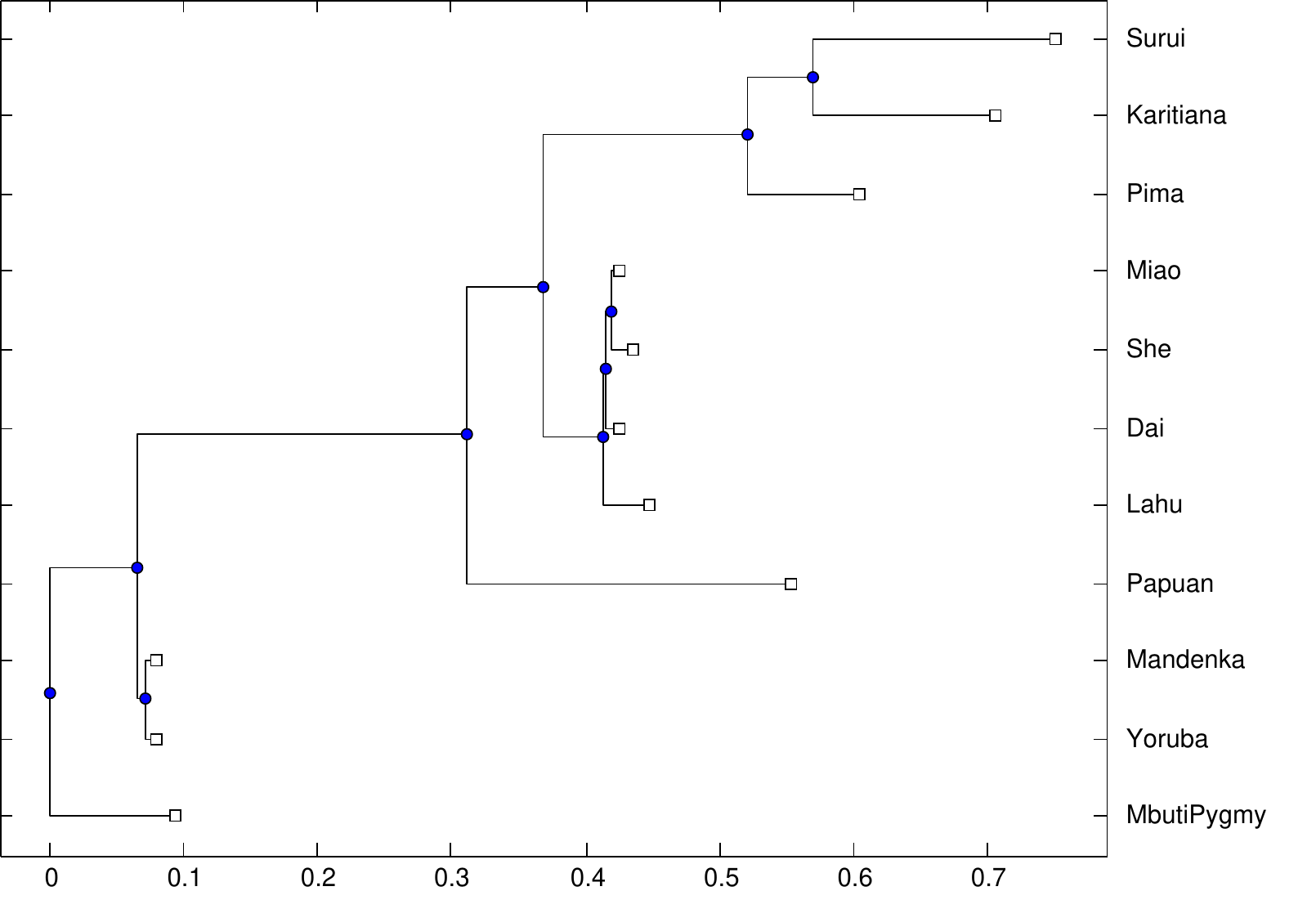}
\end{center}
\caption{{\bf Alternative scaffold tree with 11 populations used to evaluate robustness of results to scaffold choice.}  We included Mbuti Pygmy, who are known to be admixed, to help demonstrate that \MixMapper inferences are robust to deviations from additivity in the scaffold; see Tables~\ref{tab:european_mixes_alt1}--\ref{tab:double_mixes_alt1} for full results.   Distances are in drift units.}
\label{fig:alt_scaffold}
\end{figure}

\begin{figure}[H]
\begin{center}
\includegraphics[width=\textwidth]{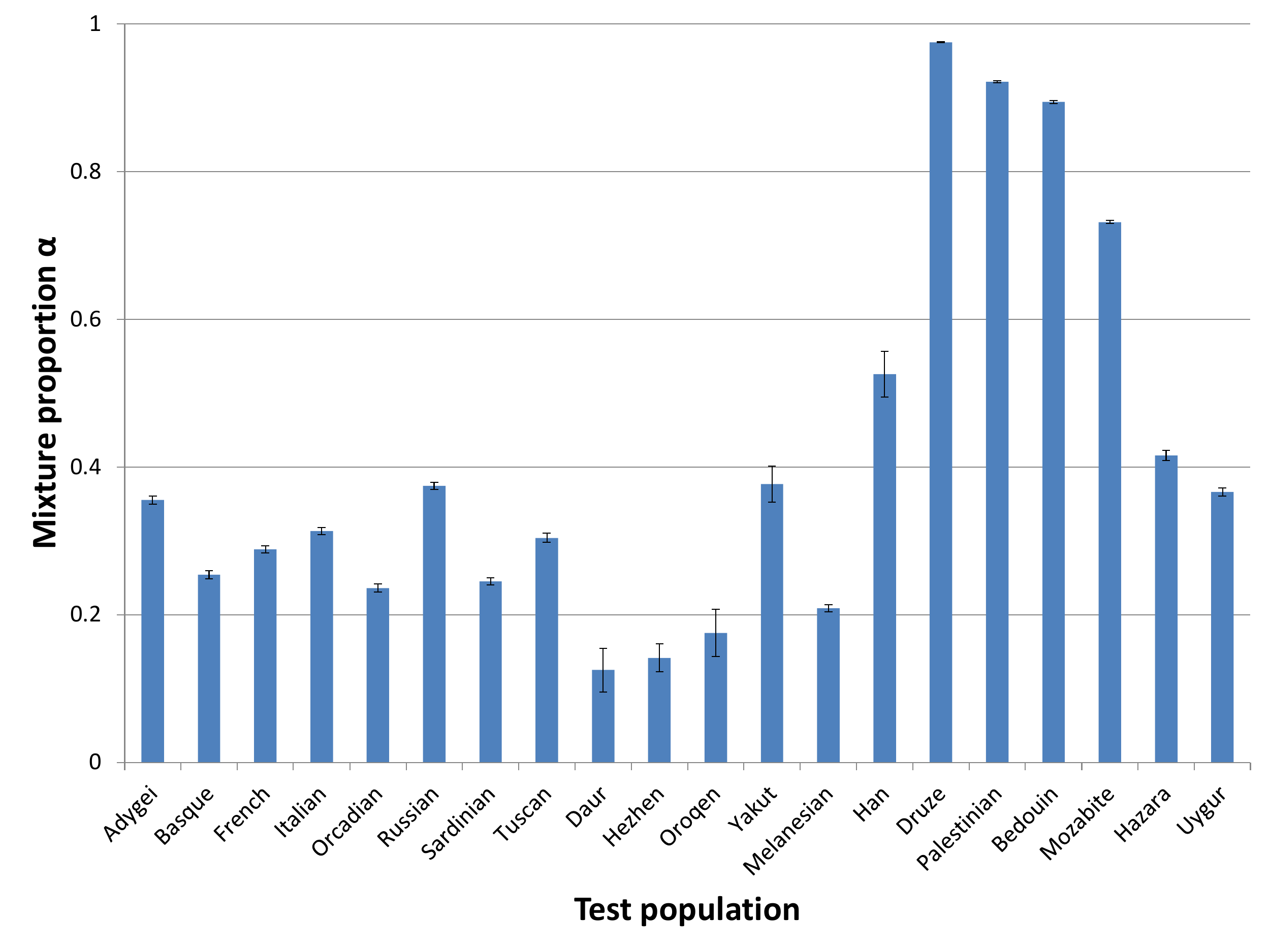}
\end{center}
\caption{{\bf Summary of mixture proportions $\alpha$ inferred with alternative 9-population scaffold trees.}  We ran \MixMapper for all 20 admixed test populations using nine different scaffold trees obtained by removing each population except Papuan one at a time from our full 10-population scaffold. (Papuan is needed to maintain continental representation.) For each test population and each scaffold, we recorded the median bootstrap-inferred value of $\alpha$ over all replicates having branching patterns similar to the primary topology.  Shown here are the means and standard deviations of the nine medians.  In all cases, $\alpha$ refers to the proportion of ancestry from the first branch as in Tables~\ref{tab:european_mixes}--\ref{tab:double_mixes}.}
\label{fig:alt_alphas}
\end{figure}

\begin{figure}[H]
\begin{center}
\includegraphics{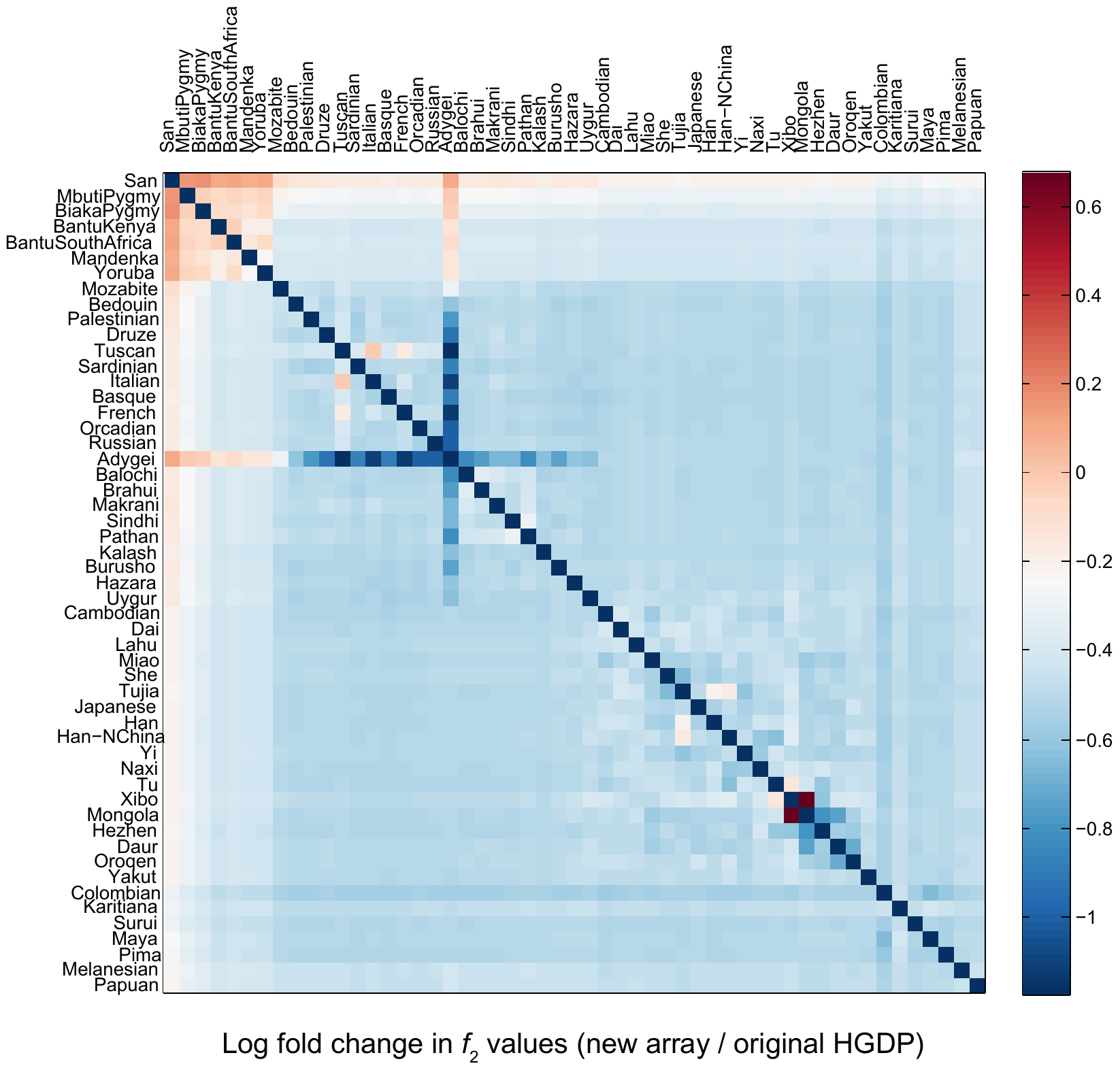}
\end{center}
\caption{{\bf Comparison of $f_2$ distances computed using original Illumina vs.\ San-ascertained SNPs.}  The heat map shows the log fold change in $f_2$ values obtained from the original HGDP data~\citep{li2008worldwide} versus the San-ascertained data~\citep{draft7} used in this study.}
\label{fig:f2_diff_heatmap}
\end{figure}

\begin{figure}[H]
\begin{center}
\includegraphics{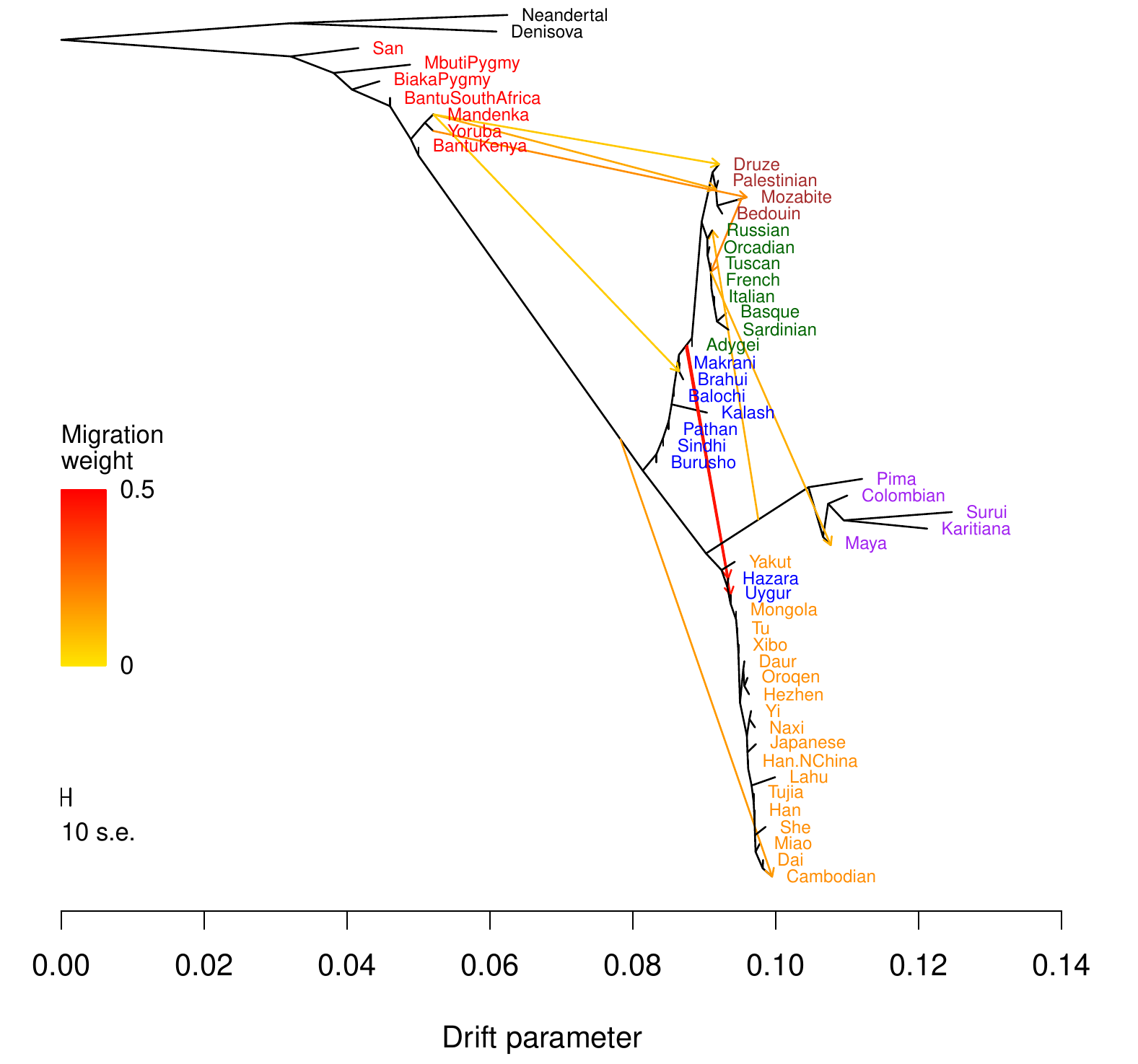}
\end{center}
\caption{{\bf \TreeMix results on the HGDP.} Admixture graph for HGDP populations obtained with the \TreeMix software, as reported in \citet{pickrell2012inference}.  Figure is reproduced from \citet{pickrell2012inference} with permission of the authors and under the Creative Commons Attribution License.}
\label{fig:TreeMix}
\end{figure}

\begin{figure}[H]
\begin{center}
\includegraphics{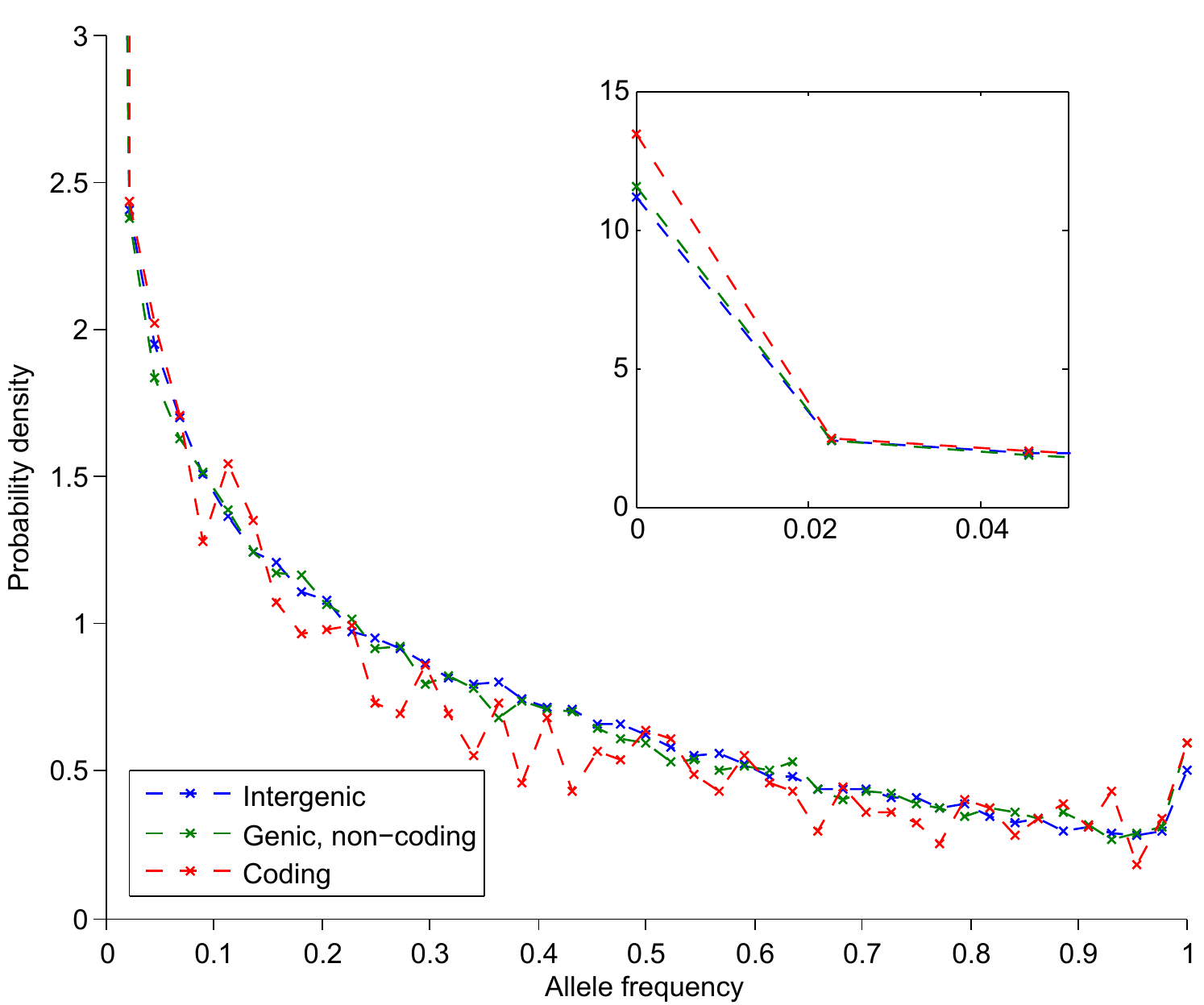}
\end{center}
\caption{{\bf Comparison of allele frequency spectra within and outside gene regions.}  We divided the Panel 4 (San-ascertained) SNPs into three groups: those outside gene regions (101,944), those within gene regions but not in exons (58,110), and those within coding regions (3259).  Allele frequency spectra restricted to each group are shown for the Yoruba population.  Reduced heterozygosity within exon regions is evident, which suggests the action of purifying selection.  (Inset) We observe the same effect in the genic, non-coding spectrum; it is less noticeable but can be seen at the edge of the spectrum.}
\label{fig:spectral_shift_near_genes}
\end{figure}

\newpage
\begin{figure}[H]
\begin{center}
\includegraphics[scale=1.5]{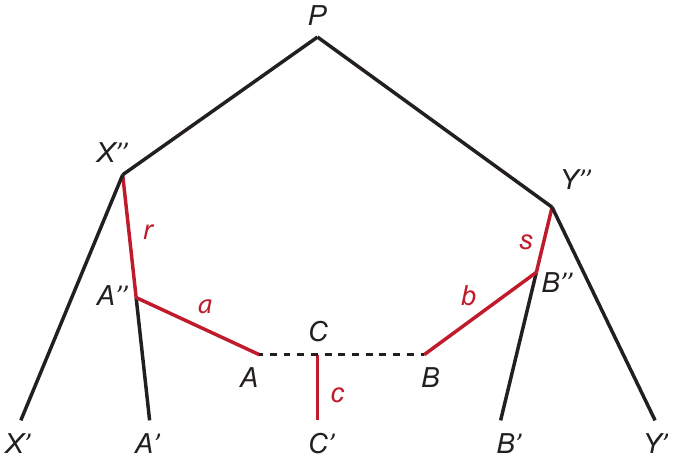}
\end{center}
\caption{{\bf Schematic of part of an admixture tree.}  Population $C$ is derived from an admixture of populations $A$ and $B$ with proportion $\alpha$ coming from $A$.  The $f_2$ distances from $C'$ to the present-day populations $A', B', X', Y'$ give four relations from which we are able to infer four parameters: the mixture fraction $\alpha$, the locations of the split points $A''$ and $B''$ (i.e., $r$ and $s$), and the combined drift $\alpha^2a + (1-\alpha)^2b + c$.  
}
\label{fig:simpleequationdiagram}
\end{figure}

\setcounter{table}{0}
\renewcommand{\thetable}{S\arabic{table}}

\newpage
\begin{table}[H]
\caption{\bf{Mixture parameters for simulated data.}}
{\scriptsize\sffamily
\begin{tabular}{l l r l l l l}
\hline
AdmixedPop       & Branch1 + Branch2    &    \# rep &  $\alpha$     &  Branch1Loc       &  Branch2Loc       &  MixedDrift \\
\hline
\bf{First tree} &&&&&&\\
pop6         & pop3 + pop5          &  500 &   0.253-0.480  &  0.078-0.195 / 0.214  &  0.050-0.086 / 0.143  &  0.056-0.068\\
pop6 (true)     & pop3 + pop5      &          &     0.4   &  0.107 / 0.213  &  0.077 / 0.145  &  0.066 \\
\hline
\bf{Second tree} &&&&&&\\
pop4      & pop3 + pop5         &  500 &   0.382-0.652  &  0.039-0.071 / 0.076  &  0.032-0.073 / 0.077  &  0.010-0.020\\
pop4 (true)     & pop3 + pop5        &   &   0.4   &   0.071 / 0.077  &   0.038 / 0.077  & 0.016 \\
pop9       & Anc3--7 + pop7     &  490 &   0.653-0.915  &  0.048-0.091 / 0.140  &  0.013-0.134 / 0.147  &  0.194-0.216\\
pop9 (true)     & Anc3--7 + pop7     &   &   0.8  &   0.077 / 0.145  &   0.037 / 0.145  &  0.194\\
pop10     & Anc3--7 + pop7     &  500 &    0.502-0.690  &  0.047-0.091 / 0.140  &  0.021-0.067 / 0.147  &  0.151-0.167\\
pop10 (true)    & Anc3--7 + pop7     &   &    0.6  &  0.077 / 0.145  &  0.037 / 0.145  &  0.150\\
\hline
AdmixedPop2   & AdmixedPop1 + Branch3  &   \# rep &  $\alpha_2$   &  Branch3Loc  &  & \\
\hline
pop8   & pop10 + pop2  &  304 &   0.782-0.822  &  0.007-0.040 / 0.040  &   &\\
   & pop10 + Anc1--2      &  193 &   0.578-0.756  &  0.009-0.104 / 0.148  &   &  \\
pop8 (true)  & pop10 + pop2  &   &   0.8  &  0.020 / 0.039  &    &    \\
\hline
\end{tabular}
}
\begin{flushleft} \textsc{Note}.---Mixture parameters inferred by \MixMapper for simulated data, followed by true values for each simulated admixed population.  Branch1 and Branch2 are the optimal split points for the mixing populations, with $\alpha$ the proportion of ancestry from Branch1; topologies are shown that that occur for at least 20 of 500 bootstrap replicates.  The mixed drift parameters for the three-way admixed pop8 are not well-defined in the simulated tree and are omitted.  The branch ``Anc3--7'' is the common ancestral branch of pops 3--7, and the branch ``Anc1--2'' is the common ancestral branch of pops 1--2. See Figure~\ref{fig:table_key} and the caption of Table~\ref{tab:european_mixes} for descriptions of the parameters and Figure~\ref{fig:simulations} for plots of the results.
\end{flushleft}
\label{tab:sim_details}
\end{table}

\newpage
\begin{table}[H]
\caption{\bf{Mixture parameters for Europeans inferred with an alternative scaffold tree.}}
{\scriptsize\sffamily
\begin{tabular}{l r l l l l}
\hline
AdmixedPop   &   \# rep &  $\alpha$   &  Branch1Loc (Anc.\ N.\ Eurasian) &  Branch2Loc (Anc.\ W.\ Eurasian) &  MixedDrift\\
\hline
Adygei           &  488 &  0.278-0.475  &  0.035-0.078 / 0.151  &  0.158-0.191 / 0.246  &  0.078-0.093\\
Basque           &  273 &  0.221-0.399  &  0.055-0.111 / 0.153  &  0.164-0.194 / 0.244  &  0.108-0.124\\
French           &  380 &  0.240-0.410  &  0.054-0.108 / 0.152  &  0.165-0.192 / 0.245  &  0.093-0.106\\
Italian          &  427 &  0.245-0.426  &  0.047-0.103 / 0.152  &  0.155-0.188 / 0.246  &  0.095-0.110\\
Orcadian         &  226 &  0.214-0.387  &  0.061-0.131 / 0.153  &  0.174-0.197 / 0.244  &  0.098-0.116\\
Russian          &  472 &  0.296-0.490  &  0.047-0.093 / 0.151  &  0.165-0.197 / 0.246  &  0.080-0.095\\
Sardinian        &  390 &  0.189-0.373  &  0.045-0.104 / 0.152  &  0.160-0.190 / 0.245  &  0.110-0.125\\
Tuscan           &  413 &  0.238-0.451  &  0.039-0.096 / 0.152  &  0.153-0.191 / 0.245  &  0.093-0.111\\
\hline
\end{tabular}
}
\begin{flushleft} \textsc{Note}.---Mixture parameters inferred by \MixMapper for modern-day European populations using an alternative unadmixed scaffold tree containing 11 populations: Yoruba, Mandenka, Mbuti Pygmy, Papuan, Dai, Lahu, Miao, She, Karitiana, Suru\'{i}, and Pima (see Figure~\ref{fig:alt_scaffold}).  
The parameter estimates are very similar to those obtained with the original scaffold tree (Table~\ref{tab:european_mixes}), with $\alpha$ slightly higher on average.  The bootstrap support for the branching position of ``ancient northern Eurasian'' plus ``ancient western Eurasian'' is also somewhat lower, with the remaining replicates almost all placing the first ancestral population along the Pima branch instead.  However, this is perhaps not surprising given evidence of European-related admixture in Pima; overall, our conclusions are unchanged, and the results appear quite robust to perturbations in the scaffold.  See Figure~\ref{fig:table_key}A and the caption of Table~\ref{tab:european_mixes} for descriptions of the parameters.
\end{flushleft}
\label{tab:european_mixes_alt1}
\end{table}

\newpage
\begin{table}[H]
\caption{\bf{Mixture parameters for other populations modeled as two-way admixtures inferred with an alternative scaffold tree.}}
{\scriptsize\sffamily
\begin{tabular}{l l r l l l l}
\hline
AdmixedPop       & Branch1 + Branch2    &    \# rep &  $\alpha$     &  Branch1Loc       &  Branch2Loc       &  MixedDrift \\
\hline
Daur       & Anc.\ N.\ Eurasian + She    &  264 &  0.225-0.459  &  0.005-0.052 / 0.151  &  0.002-0.014 / 0.016  &  0.014-0.024\\
      & Anc.\ N.\ Eurasian + Miao      &  213 &  0.235-0.422  &  0.005-0.049 / 0.151  &  0.002-0.008 / 0.008  &  0.014-0.024\\
\hline
Hezhen   & Anc.\ N.\ Eurasian + She   &  257 &  0.230-0.442  &  0.005-0.050 / 0.151  &  0.002-0.010 / 0.016  &  0.012-0.034\\
       & Anc.\ N.\ Eurasian + Miao    &  217 &  0.214-0.444  &  0.005-0.047 / 0.151  &  0.002-0.008 / 0.008  &  0.013-0.037\\
\hline
Oroqen    & Anc.\ N.\ Eurasian + She   &  336 &  0.284-0.498  &  0.010-0.052 / 0.151  &  0.003-0.015 / 0.016  &  0.017-0.036\\
         & Anc.\ N.\ Eurasian + Miao     &  149 &  0.271-0.476  &  0.007-0.046 / 0.151  &  0.002-0.008 / 0.008  &  0.018-0.039\\
\hline
Yakut    & Anc.\ N.\ Eurasian + Miao        &  246 &  0.648-0.864  &  0.004-0.018 / 0.151  &  0.005-0.008 / 0.008  &  0.032-0.043\\
          & Anc.\ East Asian + Pima       &   71 &  0.917-0.973  &  0.008-0.020 / 0.045  &  0.022-0.083 / 0.083  &  0.028-0.042\\
          & Anc.\ N.\ Eurasian + She      &  161 &  0.664-0.865  &  0.004-0.018 / 0.151  &  0.003-0.017 / 0.017  &  0.030-0.043\\
\hline
Melanesian       & Dai + Papuan           &  331 &  0.168-0.268  &  0.009-0.011 / 0.011  &  0.167-0.204 / 0.246  &  0.089-0.115\\
     & Lahu + Papuan             &   78 &  0.174-0.266  &  0.005-0.034 / 0.034  &  0.167-0.203 / 0.244  &  0.089-0.118\\
\hline
Han              & Karitiana + She        &  167 &  0.007-0.025  &  0.026-0.134 / 0.134  &  0.001-0.006 / 0.016  &  0.000-0.004\\
         & She + Surui                &   54 &  0.971-0.994  &  0.001-0.006 / 0.016  &  0.017-0.180 / 0.180  &  0.000-0.003\\
          & Anc.\ N.\ Eurasian + She       &   65 &  0.021-0.080  &  0.004-0.105 / 0.152  &  0.001-0.007 / 0.016  &  0.000-0.003\\
           & Pima + She                 &   82 &  0.009-0.033  &  0.022-0.085 / 0.085  &  0.001-0.007 / 0.016  &  0.000-0.004\\
\hline
\end{tabular}
}
\begin{flushleft} \textsc{Note}.---Mixture parameters inferred by \MixMapper for non-European populations fit as two-way admixtures using an alternative unadmixed scaffold tree containing 11 populations: Yoruba, Mandenka, Mbuti Pygmy, Papuan, Dai, Lahu, Miao, She, Karitiana, Suru\'{i}, and Pima (see Figure~\ref{fig:alt_scaffold}). The results for the first four populations are very similar to those obtained with the original scaffold tree, except that $\alpha$ is now estimated to be roughly 20\% higher.  Melanesian is fit essentially identically as before. Han, however, now appears nearly unadmixed, which we suspect is due to the lack of an appropriate northern East Asian population related to one ancestor (having removed Japanese).  See Figure~\ref{fig:table_key}A and the caption of Table~\ref{tab:european_mixes} for descriptions of the parameters; branch choices are shown that that occur for at least 50 of 500 bootstrap replicates. The ``Anc.\ East Asian'' branch is the common ancestral branch of the four East Asian populations in the unadmixed tree.
\end{flushleft}
\label{tab:other_single_mixes_alt1}
\end{table}

\newpage
\begin{table}[H]
\caption{\bf{Mixture parameters for populations modeled as three-way admixtures inferred with an alternative scaffold tree.}}
{\scriptsize\sffamily
\begin{tabular}{l l r l l l l l}
\hline
AdmixedPop2      & Branch3       &    \# rep &  $\alpha_2$   &  Branch3Loc     &  MixedDrift1A &  FinalDrift1B &  MixedDrift2\\
\hline
Druze            & Mandenka        &  309 &  0.958-0.984  &  0.004-0.009 / 0.009  &  0.088-0.102  &  0.021-0.029  &  0.005-0.013\\
\hline
Palestinian      & Mandenka       &  249 &  0.907-0.935  &  0.008-0.009 / 0.009  &  0.087-0.100  &  0.022-0.030  &  0.001-0.008\\
   & Anc.\ W.\ Eurasian          &   92 &  0.822-0.893  &  0.050-0.122 / 0.246  &  0.102-0.126  &  0.000-0.019  &  0.011-0.023\\
\hline
Bedouin          & Mandenka     &  303 &  0.852-0.918  &  0.006-0.009 / 0.009  &  0.086-0.101  &  0.022-0.030  &  0.007-0.019\\
\hline
Mozabite         & Mandenka     &  339 &  0.684-0.778  &  0.006-0.009 / 0.009  &  0.095-0.112  &  0.010-0.021  &  0.018-0.032\\
         &  Yoruba             &   50 &  0.673-0.778  &  0.005-0.010 / 0.010  &  0.093-0.111  &  0.010-0.020  &  0.018-0.031\\
\hline
Hazara     & Anc.\ East Asian  &  390 &  0.350-0.464  &  0.009-0.023 / 0.045  &  0.084-0.119  &  0.001-0.033  &  0.004-0.012\\
\hline
Uygur      & Anc.\ East Asian   &  390 &  0.312-0.432  &  0.007-0.022 / 0.045  &  0.091-0.124  &  0.000-0.027  &  0.000-0.009\\
\hline
\end{tabular}
}
\begin{flushleft} \textsc{Note}.---Mixture parameters inferred by \MixMapper for populations fit as three-way admixtures using an alternative unadmixed scaffold tree containing 11 populations: Yoruba, Mandenka, Mbuti Pygmy, Papuan, Dai, Lahu, Miao, She, Karitiana, Suru\'{i}, and Pima (see Figure~\ref{fig:alt_scaffold}).  In all cases one parent population splits from the (admixed) Sardinian branch and the other from Branch3. All the parameters are quite similar to those obtained with the original scaffold with only some relative changes in bootstrap support among alternative topologies.  See Figure~\ref{fig:table_key}B and the caption of Table~\ref{tab:european_mixes} for further descriptions of the parameters; branch choices are shown that that occur for at least 50 of the 390 bootstrap replicates having the majority branch choices for the two-way Sardinian fit.  The ``Anc.\ East Asian'' branch is the common ancestral branch of the four East Asian populations in the unadmixed tree.
\end{flushleft}
\label{tab:double_mixes_alt1}
\end{table}

\newpage
\begin{table}[H]
\caption{\bf{Mixture proportions for Sardinian and Basque from $f_4$ ratio estimation.}}
{\small\sffamily
\begin{tabular}{l l l l}
\hline
Test pop. & Asian pop. & American pop. & $\alpha$ \\
\hline
Sardinian & Dai & Karitiana & 23.3 $\pm$ 6.3 \\
Sardinian & Dai & Suru\'{i} & 24.5 $\pm$ 6.7 \\
Sardinian & Lahu & Karitiana & 23.1 $\pm$ 7.0 \\
Sardinian & Lahu & Suru\'{i} & 24.7 $\pm$ 7.6 \\
Basque & Dai & Karitiana & 22.8 $\pm$ 7.0 \\
Basque & Dai & Suru\'{i} & 24.0 $\pm$ 7.6 \\
Basque & Lahu & Karitiana & 23.1 $\pm$ 7.4 \\
Basque & Lahu & Suru\'{i} & 24.7 $\pm$ 8.0 \\
\hline
\end{tabular}
}
\begin{flushleft}
\textsc{Note}.---To validate the mixture proportions estimated by \MixMapper for Sardinian and Basque, we applied $f_4$ ratio estimation.  The fraction $\alpha$ of ``ancient northern Eurasian'' ancestry was estimated as $\alpha = f_4$(Papuan, Asian; Yoruba, European) / $f_4$(Papuan, Asian; Yoruba, American), where the European population is Sardinian or Basque, Asian is Dai or Lahu, and American is Karitiana or Suru\'{i}.  Standard errors are from 500 bootstrap replicates.  Note that this calculation assumes the topology of the ancestral mixing populations as inferred by \MixMapper (Figure~\ref{fig:european_detail}A).
\end{flushleft}
\label{tab:f4_ratio_Sardinian_Basque}
\end{table}

\setcounter{table}{0}
\renewcommand{\thetable}{S\arabic{table}}
\renewcommand{\tablename}{Text}

\newpage
\begin{table}[!ht]
  \caption{\bf{$f$-statistics and population admixture.}}
  \label{text:f-stats+admixture}
\end{table}

Here we include derivations of the allele frequency divergence equations solved by \MixMapper to determine the optimal placement of admixed populations.  These results were first presented in~\citet{India} and \citet{draft7}, and we reproduce them here for completeness, with slightly different emphasis and notation.  We also describe in the final paragraph (and in more detail in Material and Methods) how the structure of the equations leads to a particular form of the system for a full admixture tree.

Our basic quantity of interest is the $f$-statistic $f_2$, as defined in~\citet{India}, which is the squared allele frequency difference between two populations at a biallelic SNP.  That is, at SNP locus $i$, we define
\[
f_2^i(A,B) := (p_A-p_B)^2,
\]
where $p_A$ is the frequency of one allele in population $A$ and $p_B$ is the frequency of the allele in population $B$.  This is the same as Nei's minimum genetic distance $D_{AB}$ for the case of a biallelic locus~\citep{nei1987molecular}.  As in~\citet{India}, we define the unbiased %
estimator $\hat{f}_2^i(A,B)$, which is a function of finite population samples: 
\[
\hat{f}_2^i(A,B) := \left(\hat{p}_A-\hat{p}_B\right)^2-\frac{\hat{p}_A(1-\hat{p}_A)}{n_A-1}-\frac{\hat{p}_B(1-\hat{p}_B)}{n_B-1},
\]
where, for each of $A$ and $B$, $\hat{p}$ is the the empirical allele frequency and $n$ is the total number of sampled alleles.  %

We can also think of $f_2^i(A,B)$ itself as the outcome of a random process of genetic history.  In this context, we define
\[
F_2^i(A,B) := E((p_A-p_B)^2),
\]
the expectation of $(p_A-p_B)^2$ as a function of population parameters.  So, for example, if $B$ is descended from $A$ via one generation of Wright-Fisher genetic drift in a population of size $N$, then $F_2^i(A,B) = p_A(1-p_A)/2N$.

While $\hat{f}_2^i(A,B)$ is unbiased, its variance may be large, so in practice, we use the statistic
\[
\hat{f}_2(A,B) := \frac{1}{m} \sum_{i=1}^m \hat{f}_2^i(A,B),
\]
i.e., the average of $\hat{f}_2^i(A,B)$ over a set of $m$ SNPs.  As we discuss in more detail in Text~\ref{text:het+drift}, $F_2^i(A,B)$ is not the same for different loci, meaning $\hat{f}_2(A,B)$ will depend on the choice of SNPs.  However, we do know that $\hat{f}_2(A,B)$ is an unbiased estimator of the true average $f_2(A,B)$ of $f_2^i(A,B)$ over the set of SNPs.

The utility of the $f_2$ statistic is due largely to the relative ease of deriving equations for its expectation between populations on an admixture tree.  The following derivations are borrowed from~\citep{India}.  
As above, let the frequency of a SNP i in population X be $p_X$.  Then, for example,
\begin{eqnarray*}
E(f_2^i(A,B))
& = & E((p_A-p_B)^2) \\
& = & E((p_A-p_P+p_P-p_B)^2) \\
& = & E((p_A-p_P)^2) + E((p_P-p_B)^2) + 2E((p_A-p_P)(p_P-p_B)) \\
& = & E(f_2^i(A,P)) + E(f_2^i(B,P)),
\end{eqnarray*}
since the genetic drifts $p_A-p_P$ and $p_P-p_B$ are uncorrelated and have expectation 0.  We can decompose these terms further; if $Q$ is a population along the branch between $A$ and $P$, then:
\begin{eqnarray*}
E(f_2^i(A,P))
& = & E((p_A-p_P)^2) \\
& = & E((p_A-p_Q+p_Q-p_P)^2) \\
& = & E((p_A-p_Q)^2) + E((p_Q-p_P)^2) + 2E((p_A-p_Q)(p_Q-p_P)) \\
& = & E(f_2^i(A,Q)) + E(f_2^i(Q,P)).
\end{eqnarray*}
Here, again, $E(p_A-p_Q) = E(p_Q-p_P) = 0$, but $p_A-p_Q$ and $p_Q-p_P$ are not independent; for example, if $p_Q-p_P =-p_P$, i.e. $p_Q = 0$, then necessarily $p_A-p_Q = 0$.  However, $p_A-p_Q$ and $p_Q-p_P$ are independent conditional on a single value of $p_Q$, meaning the conditional expectation of $(p_A-p_Q)(p_Q-p_P)$ is 0.  By the double expectation theorem,
\[
E((p_A-p_Q)(p_Q-p_P)) = E(E((p_A-p_Q)(p_Q-p_P)|p_Q)) = E(E(0)) = 0.
\]
From $E(f_2^i(A,P)) = E(f_2^i(A,Q)) + E(f_2^i(Q,P))$, we can take the average over a set of SNPs to yield, in the notation from above,
\[
F_2(A,P) = F_2(A,Q) + F_2(Q,P).
\]

We have thus shown that $f_2$ distances are additive along an unadmixed-drift tree.  This property is fundamental for our theoretical results and is also essential for finding admixtures, since, as we will see, additivity does not hold for admixed populations.

Given a set of populations with allele frequencies at a set of SNPs, we can use the estimator $\hat{f}_2$ to compute $f_2$ distances between each pair.  These distances should be additive if the populations are related as a true tree.  Thus, it is natural to build a phylogeny using neighbor-joining~\citep{saitou1987neighbor}, yielding a fully parameterized tree with all branch lengths inferred.   However, in practice, the tree will not exactly be additive, and we may wish to try fitting some population $C'$ as an admixture.  To do so, we would have to specify six parameters (in the notation of Figure~\ref{fig:simpleequationdiagram}): the locations on the tree of $A''$ and $B''$; the branch lengths $f_2(A'',A)$, $f_2(B'',B)$, and $f_2(C,C')$; and the mixture fraction.  These are the variables $r$, $s$, $a$, $b$, $c$, and $\alpha$.

In order to fit $C'$ onto an unadmixed tree (that is, solve for the six mixture parameters), we use the equations for the expectations $F_2(C',Z')$ of the $f_2$ distances between $C'$ and each other population $Z'$ in the tree.  Referring to Figure~\ref{fig:simpleequationdiagram}, with the point admixture model, the allele frequency in $C$ is $p_C = \alpha~p_A+(1-\alpha)~p_B$.  So, for a single locus, using additivity,
\begin{eqnarray*}
E(f_2^i(A',C'))
& = & E((p_{A'}-p_{C'})^2) \\
& = & E((p_{A'}-p_{A''}+p_{A''}-p_C+p_C-p_{C'})^2) \\
& = & E((p_{A'}-p_{A''})^2) + E((p_{A''}-\alpha~p_A-(1-\alpha)~p_B)^2) + E((p_C-p_{C'})^2) \\
& = & E(f_2^i(A',A'')) + \alpha^2E(f_2^i(A'',A)) \\ && + (1-\alpha)^2E(f_2^i(A'',B)) + E(f_2^i(C,C')).  
\end{eqnarray*}
Averaging over SNPs, and replacing $E(f_2(A',C'))$ by the estimator $\hat{f}_2(A',C')$, this becomes 
\begin{eqnarray*}
\hat{f}_2(A',C') & = & F_2(A',X'') - r + \alpha^2a \\ && + (1-\alpha)^2(r + F_2(X'',Y'') + s + b) + c\\
\implies \hat{f}_2(A',C') - F_2(A',X'') & = & (\alpha^2 - 2\alpha)r + (1-\alpha)^2s + \alpha^2a \\ && + (1-\alpha)^2b + c + (1-\alpha)^2F_2(X'',Y'').
\end{eqnarray*}
The quantities $F_2(X'',Y'')$ and $F_2(A',X'')$ are constants that can be read off of the neighbor-joining tree.  Similarly, we have 
\begin{equation*}
\hat f_2(B',C') - F_2(B',Y'') = \alpha^2 r + (\alpha^2 - 1) s + \alpha^2a+
(1-\alpha)^2 b + c + \alpha^2 F_2(X'',Y'').
\end{equation*}
For the outgroups $X'$ and $Y'$, we have
\begin{eqnarray*}
\hat f_2(X',C') &=& \alpha^2(c + a + r + F_2(X',X''))
\\&&+(1-\alpha)^2(c + b + s + F_2(X'',Y'') +F_2(X',X''))
\\&&+2\alpha(1-\alpha) \left(c + F_2(X',X'')\right)\\
&=& \alpha^2 r +(1-\alpha)^2 s + \alpha^2a+(1-\alpha)^2 b + c 
\\&& + (1-\alpha)^2 F_2(X'',Y'') +F_2(X',X'')
\end{eqnarray*}
and 
\begin{equation*}
\hat f_2(Y',C') = \alpha^2 r  + (1-\alpha)^2 s + \alpha^2 a + (1-\alpha)^2 b +c
+ \alpha^2 F_2(X'',Y'')+ F_2(Y',Y'').
\end{equation*}

Assuming additivity within the neighbor-joining tree, any population descended from $A''$ will give the same equation (the first type), as will any population descended from $B''$ (the second type), and any outgroup (the third type, up to a constant and a coefficient of $\alpha$).  Thus, no matter how many populations there are in the unadmixed tree---and assuming there are at least two outgroups $X'$ and $Y'$ such that the points $X''$ and $Y''$ are distinct---the system of equations consisting of $E(f_2(P,C'))$ for all $P$ will contain precisely enough information to solve for $\alpha$, $r$, $s$, and the linear combination $\alpha^2a + (1-\alpha)^2b + c$.  We also note the useful fact that for a fixed value of $\alpha$, the system is linear in the remaining variables.

\newpage
\begin{table}[!ht]
  \caption{\bf{Heterozygosity and drift lengths.}}
  \label{text:het+drift}
\end{table}

One disadvantage to building trees with $f_2$ statistics is that the values are not in easily interpretable units.  For a single locus, the $f_2$ statistic measures the squared allele frequency change between two populations.  However, in practice, one needs to compute an average $f_2$ value over many loci.  Since the amount of drift per generation is proportional to $p(1-p)$, the expected frequency change in a given time interval will be different for loci with different initial frequencies.  This means that the estimator $\hat{f}_2$ depends on the distribution of frequencies of the SNPs used to calculate it.  For example, within an $f_2$-based phylogeny, the lengths of non-adjacent edges are not directly comparable.

In order to make use of the properties of $f_2$ statistics for admixture tree building and still be able to present our final trees in more directly meaningful units, we will show now how $f_2$ distances can be converted into absolute drift lengths. %
Again, we consider a biallelic, neutral SNP in two populations, with no further mutations, under a Wright-Fisher model of genetic drift. 

Suppose populations $A$ and $B$ are descended independently from a population $P$, and we have an allele with frequency $p$ in $P$, $p_A=p+a$ in $A$, and $p_B=p+b$ in $B$.  The (true) heterozygosities at this locus are $h_P^i = 2p(1-p)$, $h_A^i = 2p_A(1-p_A)$, and $h_B^i = 2p_B(1-p_B)$. As above, we write $\hat{h}^i_A$ for the unbiased single-locus estimator
\[
\hat{h}^i_A := \frac{2n_A\hat{p}_A(1-\hat{p}_A)}{n_A-1},
\]
$\hat{h}_A$ for the multi-locus average of $\hat{h}^i_A$, and $H_A^i$ for the expectation of $h_A^i$ under the Wright-Fisher model (and similarly for $B$ and $P$).

Say $A$ has experienced $t_A$ generations of drift with effective population size $N_A$ since the split from $P$, and $B$ has experienced $t_B$ generations of drift with effective population size $N_B$.  Then it is well known that $H_A^i = h_P^i(1-D_A)$, where $D_A = 1 - (1-1/(2N_A))^{t_A}$, and $H_B^i = h_P^i(1-D_B)$.  We also have
\begin{eqnarray*}
H_A^i
& = & E(2(p+a)(1-p-a)) \\
& = & E(h_P^i-2ap+2a-2ap-2a^2) \\
& = & h_P^i-2E(a^2) \\
& = & h_P^i-2F_2^i(A, P),
\end{eqnarray*}
so $2F_2^i(A, P) = h_P^iD_A$.  Likewise, $2F_2^i(B, P) = h_P^iD_B$ and $2F_2^i(A, B) = h_P^i(D_A+D_B)$.  Finally,
\[
H_A^i+H_B^i+2F_2^i(A, B) = h_P^i(1-D_A) + h_P^i(1-D_B) + h_P^i(D_A+D_B) = 2h_P^i.
\]
This equation is essentially equivalent to one in~\citet{nei1987molecular}, although Nei interprets his version as a way to calculate the expected present-day heterozygosity rather than estimate the ancestral heterozygosity.  To our knowledge, the equation has not been applied in the past for this second purpose.

In terms of allele frequencies, the form of $h_P^i$ turns out to be very simple:
\[
h_P^i = p_A+p_B-2p_Ap_B = p_A(1-p_B)+p_B(1-p_A),
\]
which is the probability that two alleles, one sampled from $A$ and one from $B$, are different by state.  We can see, therefore, that this probability remains constant in expectation after any amount of drift in $A$ and $B$.  This fact is easily proved directly:
\[
E(p_A+p_B-2p_Ap_B) = 2p-2p^2 = h_P^i,
\]
where we use the independence of drift in $A$ and $B$.

Let $\hat{h}_P^i := (\hat{h}_A^i+\hat{h}_B^i+2\hat{f}_2^i(A, B))/2$, and let $h_P$ denote the true average heterozygosity in $P$ over an entire set of SNPs.  Since $\hat{h}^i_P$ is an unbiased estimator of $(h_A^i+h_B^i+2f_2^i(A, B))/2$, its expectation under the Wright-Fisher model is $h_P^i$.  So, the average $\hat{h}_P$ of $\hat{h}_P^i$ over a set of SNPs is an unbiased (and potentially low-variance) estimator of $h_P$.  If we have already constructed a phylogenetic tree using pairwise $f_2$ statistics, we can use the inferred branch length $\hat{f}_2(A', P)$ from a present-day population $A$ to an ancestor $P$ in order to estimate $\hat{h}_P$ more directly as $\hat{h}_P = \hat{h}_A+2\hat{f}_2(A, P)$.  This allows us, for example, to estimate heterozygosities at intermediate points along branches or in the ancestors of present-day admixed populations.

The statistic $\hat{h}_P$ is interesting in its own right, as it gives an unbiased estimate of the heterozygosity in the common ancestor of any pair of populations (for a certain subset of the genome).  For our purposes, though, it is most useful because we can form the quotient
\[
\hat{d}_A := \frac{2\hat{f}_2(A,P)}{\hat{h}_P},
\]
where the $f_2$ statistic is inferred from a tree.  This statistic $\hat{d}_A$ is not exactly unbiased, but by the law of large numbers, if we use many SNPs, its expectation is very nearly
\[
E(\hat{d}_A) \approx \frac{E(2\hat{f}_2(A,P))}{E(\hat{h}_P)} = \frac{h_PD_A}{h_P} = D_A,
\]
where we use the fact that $D_A$ is the same for all loci.  Thus $\hat{d}$ is a simple, direct, nearly unbiased moment estimator for the drift length between a population and one of its ancestors.  This allows us to convert branch lengths from $f_2$ distances into absolute drift lengths, one branch at a time, by inferring ancestral heterozygosities and then dividing. %

For a terminal admixed branch leading to a present-day population $C'$
with heterozygosity $\hat{h}_{C'}$, we divide twice the inferred mixed
drift $c_1 = \alpha^2a+(1-\alpha)^2b+c$ (Figure~\ref{fig:table_key})
by the heterozygosity $\hat{h}_{C'}^* := \hat{h}_{C'}+2c_1$.  This is
only an approximate conversion, since it utilizes a common value
$\hat{h}_{C'}^*$ for what are really three disjoint branches, but the
error should be very small with short drifts.

An alternative definition of $\hat{d}_A$ would be $1-\hat{h}_A/\hat{h}_P$, which also has expectation (roughly) $D_A$.  In most cases, we prefer to use the definition in the previous paragraph, which allows us to leverage the greater robustness of the $f_2$ statistics, especially when taken from a multi-population tree.  %

We note that this estimate of drift lengths is similar in spirit to the widely-used statistic $F_{ST}$.  For example, under proper conditions, the expectation of $F_{ST}$ among populations that have diverged under unadmixed drift is also $1-(1-1/(2N_e))^t$~\citep{nei1987molecular}.  When $F_{ST}$ is calculated for two populations at a biallelic locus using the formula $(\Pi_D-\Pi_S)/\Pi_D$, where $\Pi_D$ is the probability two alleles from different populations are different by state and $\Pi_S$ is the (average) probability two alleles from the same population are different by state (as in~\citet{India} or the measure $G'_{ST}$ in~\citet{nei1987molecular}), then this $F_{ST}$ is exactly half of our $\hat{d}$.  As a general rule, drift lengths $\hat{d}$ are approximately twice as large as values of $F_{ST}$ reported elsewhere.

\end{document}